\documentclass[%
 aps,
 reprint,
 superscriptaddress,
 amsmath,amssymb,
 floatfix,
]{revtex4-2}

\usepackage{amsmath,amssymb,amsfonts}%
\usepackage{amsthm}%
\usepackage{mathrsfs}%
\usepackage[title]{appendix}%
\usepackage{xcolor}%
\usepackage{textcomp}%
\usepackage{booktabs}%
\usepackage{algorithm}%
\usepackage{algorithmicx}%
\usepackage{algpseudocode}%
\usepackage{listings}%
\usepackage{siunitx} 
\DeclareSIUnit{\belmilliwatt}{Bm}
\DeclareSIUnit{\dBm}{\deci\belmilliwatt}
\usepackage{braket}
\usepackage{placeins}
\usepackage[T1]{fontenc}

\usepackage{graphicx}
\usepackage{dcolumn}
\usepackage{bm}
\usepackage{bbold}

\usepackage{hyperref}
\hypersetup{
    colorlinks=true,
    linkcolor=magenta,
    citecolor=magenta,
    filecolor=magenta,      
    urlcolor=magenta,
}

\begin{document}

\title{Baseband control of single-electron silicon spin qubits in two dimensions}

\author{Florian~K.~Unseld}
\thanks{These authors contributed equally to this work.}
\author{Brennan~Undseth}
\thanks{These authors contributed equally to this work.}
\author{Eline~Raymenants}
\author{Yuta~Matsumoto}
\affiliation{QuTech and Kavli Institute of Nanoscience, Delft University of Technology, Lorentzweg 1, 2628 CJ Delft, The Netherlands}
\author{Saurabh~Karwal}
\affiliation{QuTech and Netherlands Organization for Applied Scientific Research (TNO), Stieltjesweg 1, 2628 CK Delft, Netherlands}
\author{Oriol~Pietx-Casas}
\author{Alexander~S.~Ivlev}
\author{Marcel~Meyer}
\affiliation{QuTech and Kavli Institute of Nanoscience, Delft University of Technology, Lorentzweg 1, 2628 CJ Delft, The Netherlands}
\author{Amir~Sammak}
\affiliation{QuTech and Netherlands Organization for Applied Scientific Research (TNO), Stieltjesweg 1, 2628 CK Delft, Netherlands}
\author{Menno~Veldhorst}
\author{Giordano~Scappucci}
\author{Lieven~M.~K.~Vandersypen}
\thanks{L.M.K.Vandersypen@tudelft.nl}
\affiliation{QuTech and Kavli Institute of Nanoscience, Delft University of Technology, Lorentzweg 1, 2628 CJ Delft, The Netherlands}

\begin{abstract}

Micromagnet-enabled electric-dipole spin resonance (EDSR) is an established method of high-fidelity single-spin control in silicon. However, the resulting architectural limitations have restrained silicon quantum processors to one-dimensional arrays, and heating effects from the associated microwave dissipation exacerbates crosstalk during multi-qubit operations. In contrast, qubit control based on hopping spins has recently emerged as a compelling primitive for high-fidelity baseband control in sparse two-dimensional hole arrays in germanium. In this work, we commission a $^{28}$Si/SiGe 2x2 quantum dot array both as a four-qubit device with pairwise exchange interactions using established EDSR techniques and as a two-qubit device using baseband hopping control. In this manner, we can evaluate the two modes of operation in terms of fidelity, coherence, and crosstalk. We establish a lower bound on the fidelity of the hopping gate of \SI{99.50(6)}{\percent}, which is similar to the average fidelity of the resonant gate of \SI{99.54(4)}{\percent}. Lowering the external field to reach the hopping regime nearly doubles the measured $T_2^{\mathrm{H}}$, suggesting a reduced coupling to charge noise. Finally, the hopping gate circumvents the transient pulse-induced resonance shift. To further motivate the hopping gate approach as an attractive means of scaling silicon spin-qubit arrays, we propose an extensible nanomagnet design that enables engineered baseband control of large spin arrays.

\end{abstract}

\maketitle

\section{Introduction}
Industrial fabrication compatibility is a flagship argument for semiconductor spin qubits in gate-defined quantum dots as a candidate for large-scale quantum computation and simulation \cite{Maurand_2016, Zwerver2022, Neyens2024, steinacker2024, huckemann2024}, but this promise can only be fully realized if the spin physics utilized to control qubits is also extensible. Currently, on-chip micromagnets have enabled state-of-the-art devices to exhibit powerful primitives including high-fidelity universal gate sets \cite{Xue2022, Noiri2022, Mills2022}, high-fidelity initialization and readout of multiple qubits \cite{Philips2022, Takeda_2024}, and coherent spin shuttling \cite{Noiri_2022, DeSmet2024}.

The micromagnet typically serves two purposes. It produces a longitudinal gradient parallel to the quantization axes to give each spin a unique frequency and a transversal gradient that allows microwave electric fields to drive single-qubit gates via electric-dipole spin resonance (EDSR) \cite{Neumann2015, Yoneda2015, Stuyck2021}. While this method of operation has been extended to 12 qubits \cite{george2024}, the linear charge-noise sweet spot of present magnet designs has motivated scaling in only one dimension. The limited connectivity and stringent fault tolerance thresholds of linear arrays imply that entering the second dimension is all but essential \cite{Jones2018}. Although proposals exist for scaling EDSR control into two-dimensional arrays using on-chip magnets \cite{Tadokoro2021, Aldeghi2024}, an experimental demonstration beyond 1D has, until now, been lacking.

Even in 1D arrays the dissipation of microwaves used for EDSR control is known to produce heating effects that complicate multi-qubit operation \cite{Freer_2017, Takeda2018, Undseth2023, Tanttu2024}. Typically, this manifests as qubit frequency shifts that are contextual upon the magnitude and duration of preceding microwave bursts. Circumventing this issue by working at warmer device temperatures is possible, but microwaves may be bypassed altogether. Whereas singlet-triplet and exchange-only encoded qubits allow for universal baseband control, aspects like state leakage and gate complexity have limited error rates to higher levels than for the single-spin encoding. Recently, single-hole spins in a 2D germanium array have been manipulated with high fidelity using baseband hopping control exploiting large differences in the quantization axis arising from dot-to-dot variations in the $g$-tensor \cite{van_Riggelen_Doelman_2024,Wang2024}. It is therefore intriguing whether such control can be engineered with on-chip magnets and, if so, what implications this poses for future architectures.

Here, we commission a 2x2 silicon quantum dot device for qubit control in two distinct regimes. First, we demonstrate that conventional micromagnet-based EDSR control of all four spins is possible in the 2D array, and we also demonstrate nearest-neighbor exchange control. We benchmark the performance of the four-qubit system with a particular emphasis on the crosstalk caused by off-resonant driving. Next, we lower the external magnetic field of the half-filled array to demonstrate qubit operation via hopping gates, whereby the tip in quantization axis is induced by the engineered magnetic stray field. We further analyze the qubit coherence, fidelity, and crosstalk properties of the hopping gate. Finally, we propose how a repeated nanomagnet pattern can be used to engineer hopping control across an arbitrarily large 2D array, thereby illustrating how baseband control of single-electron spins in silicon may be a compelling control method for future devices.

\begin{figure*}
    \centering
    \includegraphics[width=\textwidth]{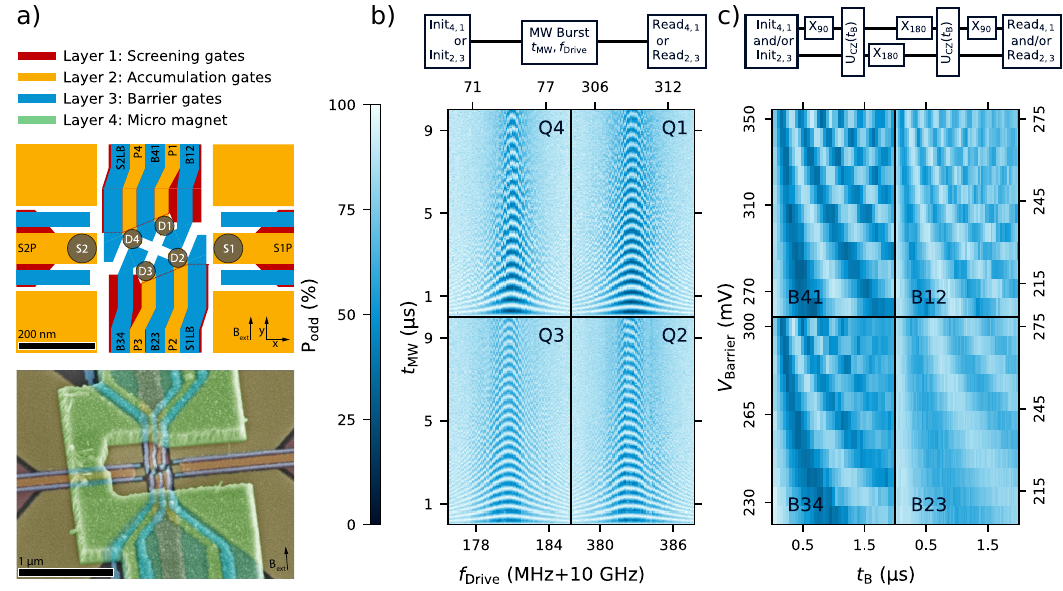}
    \caption{\textbf{A 2x2 Silicon Spin Qubit Array} a) Design schematic and false coloured SEM image of a device nominally identical to the one measured. In the schematic, single-electron transistor (SET) sensor positions are marked with S1 and S2, and the quantum dots are labeled clockwise with D1 through D4. We refer to each qubit as Q1 through Q4 according to the dot it primarily inhabits. The SEM image includes the micro magnet on top of the gate stack (not shown in the schematic). 
    b) Odd parity probability after driving the qubit with a microwave burst as indicated by the block diagram. The respective Chevron patterns of all qubits are arranged according to their physical position. The drive power and modulation amplitude were adjusted per qubit to achieve a Rabi frequency around \SI{2}{\mega\hertz}.
    c) Decoupled controlled phase oscillations of adjacent qubit pairs using the pulse sequence indicated in the block diagram. The Ising-like interaction $U_{\rm CZ}$ is controlled by the voltage $V_\mathrm{Barrier}$ on the barrier gate located between the respective plunger gates. We chose a static operating point such that the extrapolated exchange strength $J_\mathrm{off}<20$~kHz and apply a pulse on the barriers large enough such that $J_\mathrm{on}>1$~MHz. The large amplitudes for these barrier pulses as well as the particular fanout of gates B34, P3, and B23 often caused substantial degradation of the readout signal (see Section~\ref{sec:Methods}).}
    \label{Fig:Dev_Chev_DCZ}
\end{figure*}

\section{A 2x2 silicon quantum processor based on EDSR}

The most straightforward progression from a one-dimensional silicon spin-qubit array to a two-dimensional array is the adoption of existing control strategies with only minor accommodations \cite{Philips2022}. A 2x2 quantum dot array as shown in Fig.~\ref{Fig:Dev_Chev_DCZ}a) was fabricated on a $^{28}$Si/SiGe heterostructure (residual nuclear spins of 800~ppm) and is used to accumulate four single electrons under each plunger gate as has been previously demonstrated \cite{Unseld2023}. After magnetizing the micromagnet at a field of \SI{+2}{\tesla} along the positive y-axis as indicated in the SEM and schematic of Fig.~\ref{Fig:Dev_Chev_DCZ}a), the external field is reduced to \SI{200}{\milli\tesla}. We use Pauli-spin blockade (PSB) for initialization and measurement on the horizontal pairs (Q$_1$Q$_4$ and Q$_2$Q$_3$) along with electric-dipole spin resonance (EDSR) to perform addressable single-qubit rotations. The 22.5$^\circ$ rotation of the 2x2 array combined with the longitudinal field gradient gives good spectral separation of the spins with a minimum frequency difference of \SI{74}{\mega\hertz}. Using adequate parameters we find reasonable agreement with micromagnet simulations (see Extended Data Fig.~\ref{Ext_Fig:micromagnet} for details). The Rabi frequency of all four spins was tuned to \SI{2}{\mega\hertz} to stay within the linear amplitude scaling regime (see Extended Data Fig.~\ref{Ext_Fig:EDSR_Driving}) to observe the Chevron patterns in Fig.~\ref{Fig:Dev_Chev_DCZ} b).

Despite the qubits not lying along the decoherence sweet axis of the micromagnet, we observe coherence metrics similar to \cite{Philips2022}. The measured Ramsey decay times are $T_2^*=\{3.31(9), 2.03(2), 3.57(8), 2.90(5)\}~~\SI{}{\micro\second}$ for qubits Q$_1$ through Q$_4$ respectively with integration times of around \SI{30}{\minute}. Similarly, we measured Hahn-echo decay times of $T_2^{\mathrm{H}}=\{30.21(24), 16.22(18), 40.38(21), 21.77(38)\}~\SI{}{\micro\second}$. Notably, the trend of coherence times is not strictly correlated to the designed decoherence gradients at each dot location (see Extended Data Fig.~\ref{Ext_Fig:micromagnet}). There are three possible reasons for such discrepancies. First, the decoherence gradient is not isotropic in the plane of the quantum well, and the observed coherence times will depend on the microscopic orientation of the local charge noise fluctuators for each qubit. Second, large variations in coherence may occur for sufficiently sparse charge noise baths depending on the nature of the constituent fluctuators \cite{Mehmandoost_2024}. Third, hyperfine noise due to residual nuclear spins in the heterostructure may also contribute to decoherence that does not couple to the spins via the micromagnet gradient \cite{cvitkovich2024coherencelimithyperfineinteraction}.

Two-qubit interactions are controlled using the barrier gates located between neighboring plunger gate pairs to modulate the Ising-like exchange interaction. We observe the characteristic CPhase oscillations for all four nearest-neighbor pairs of qubits as a function of time and barrier pulse height as depicted in Fig.~\ref{Fig:Dev_Chev_DCZ} c). Although we operate at the symmetry point to minimize the sensitivity to charge noise, the quality factor $T_2^{\mathrm{DCZ}}/t_\mathrm{CZ}$ of the oscillations for all pairs is limited to $Q_{2Q} =\{17(2), 7.5(7), 7.1(4), 12.2(7)\}$ for B12, B23, B34, and B41 respectively, which bounds the achievable two-qubit gate fidelities to modest values. Extended Data Fig.~\ref{Ext_Fig:QST} shows state tomography results before and after applying a calibrated CZ gate for qubit pairs Q$_4$Q$_1$ and Q$_2$Q$_3$, highlighting the universal capability of this processor.

One reason for the suboptimal two-qubit performance may be the magnitude of pulses, ranging from \SI{275}{\milli\volt} to \SI{350}{\milli\volt}, required to achieve a satisfactory on/off exchange ratio as a result of the small barrier lever arms. Such large pulses jeopardize device stability and, in some cases, degrade readout visibility (see Fig.~\ref{Ext_Fig:Set_Up}). Although we apply large pulse amplitudes, the maximum achievable exchange couplings of \SIrange[range-phrase = --]{1}{4}{\mega\hertz} are small with respect to the dephasing times of the qubits.

\begin{figure*}
    \centering
    \includegraphics[width=\textwidth]{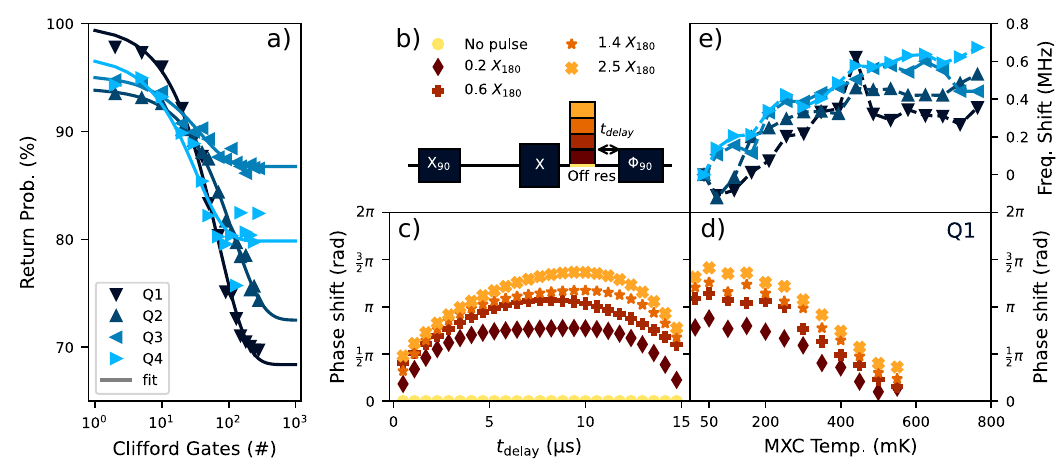}
    \caption{\textbf{EDSR Operation and PIRS Effects} 
    a) Return probability as function of the number of applied Clifford gates averaged over twenty different gate sequences for all four qubits. As each qubit readout has a different visibility, the curves decay to different values. The fitted Clifford gate fidelities are $F_\mathrm{Clif}^\mathrm{res} = \{\SI{99.28(5)}{\percent}, \SI{99.58(4)}{\percent}, \SI{98.7(2)}{\percent}, \SI{98.5(3)}{\percent}\}$ for Q$_1$ through Q$_4$ respectively. The Clifford gate set is compiled using only X$_{\pm 90}$ and Y$_{\pm 90}$ primitive gates (see Methods), and the corresponding average physical gate fidelities are, respectively, $F_\mathrm{avg}^\mathrm{res} = \{\SI{99.67(2)}{\percent}, \SI{99.80(2)}{\percent}, \SI{99.39(8)}{\percent}, \SI{99.31(15)}{\percent}\}$.
    b) Adapted Hahn Echo sequence to measure crosstalk effects as introduced in \cite{Takeda2018}. The off-resonant burst simulates the drive of a different qubit. By substituting the last X$_{90}$ with a $\Phi_{90}=\mathrm{X}_{90}\mathrm{Z}(\Phi)$ operation in the echo sequence and sweeping the phase $\Phi$, the relative phase pickup due to the off-resonant burst can be inferred. The transient behaviour is detected by varying the temporal position $t_{\mathrm{delay}}$ of the off-resonant burst with respect to $\Phi_{90}$. 
    c) Example of phase pickup as a function of $t_{\mathrm{delay}}$ for Q$_1$ at \SI{100}{\milli\kelvin} for different amplitudes of the off-resonant burst (see panel b) for legend). We use a standard Hahn-echo sequence (yellow data points) as a reference to remove constant artifacts introduced by the echo pulse sequence.
    d) The maximum phase shift extracted from measurements as in panel c) as a function of mixing chamber temperature for qubit Q$_1$ for four different off-resonance bursts (see panel b) for legend). All qubits exhibit similar behaviour (see Extended Data Fig.~\ref{Ext_Fig:EDSR_PIRS}).
    e) Temperature dependence of the bare qubit frequencies as measured by a Ramsey experiment relative to the frequency at base temperature. Symbols refer to different qubits as in panel a)}
    \label{Fig:EDSR}
\end{figure*}

Despite this shortcoming, we can still draw valuable insights about multi-qubit control with such an EDSR-based architecture. The quality factors $T_2^\mathrm{Rabi}/t_\mathrm{180}$ of the four qubits are measured to be $Q_{1Q} =\{50(2), 54(3), 55(3), 49(3)\}$, and we estimate the resonant X$_{90}$ fidelity averaged across all four qubits to be $99.54(4)\%$ using randomized benchmarking as displayed in Fig.~\ref{Fig:EDSR}a). Having established a good single-qubit gate set for each qubit, we can probe the crosstalk such gates impart on neighboring qubits. Controlling crosstalk is critical for multi-qubit operation, and while the absolute amount of crosstalk is relevant for calibration, its contextuality in time and number of gates is particularly indicative of how difficult this calibration becomes in practice. For EDSR, crosstalk commonly manifests as a spurious pulse-induced resonance shift (PIRS). We probe this ``heating effect'' on idling qubits using a modified Hahn echo sequence as depicted in Fig.~\ref{Fig:EDSR}b) \cite{Takeda2018}. The sequence includes an off-resonant microwave burst that simulates the drive of another qubit, and we quantify the magnitude of this burst in relation to the average power required to perform single-qubit rotations across all four qubits in this device. 

A nontrivial transient phase pickup is observed for all qubits for off-resonant bursts that are energetically comparable to those used for single-qubit gates. A particular instance is shown in Fig.~\ref{Fig:EDSR}c), though all qubits exhibit the same qualitative behavior (see Extended Data). With increasing delay time and amplitude, more phase is picked up until an apparent saturation is reached. This transient response has been documented in previous studies \cite{Takeda2018}. However, when the off-resonant burst occurs closer in time to the resonant echo pulse, less phase is picked up. The most likely origin of this counterintuitive behavior is the transient heating caused by the resonant echo pulse: the heating induced by the decoupling pulse reduces the transient effect of the off-resonant burst. The nonlinear combination of the two transients has severe consequences for crosstalk calibration: the required compensation will be sensitive to both pulse scheduling and amplitude.

It has previously been observed that PIRS is dependent on the temperature of the device \cite{Undseth2023}. To verify this, we repeat the above experiment on all qubits for different mixing chamber temperatures. From the fitted data (e.g. Fig.~\ref{Fig:EDSR}c)) we extract and compare the maximum phase pickup as a function of temperature.  Fig. \ref{Fig:EDSR}d) shows how increasing the mixing chamber temperature reduces the magnitude of the PIRS effect for Q1 (see Extended Data Fig.~\ref{Ext_Fig:EDSR_PIRS} for Q2-Q4). This substantial temperature increase, however, also has an impact on qubit coherence. Although the $T_2^*$ of all qubits is relatively unaffected by temperature, the $T_2^\mathrm{H}$ generally decreases monotonically as the mixing chamber temperature is raised (see Extended Data Fig.~\ref{Ext_Fig:T2vTemp}). Additionally, an increase in spin Larmor frequency as a function of temperature is observed for all quantum dots as seen in Fig.~\ref{Fig:EDSR}e).

While these results are qualitatively in agreement with a previous study of a six-spin array \cite{Undseth2023}, there are three notable differences. First, the generally monotonic increase in frequency of all four qubits studied here contrasts with the previous measurement of a non-monotonic trend across the six-qubit array. Second, the heating effect is more moderate than the \SI{1}{\mega\hertz} frequency shifts observed by Undseth et al. when using similar microwave bursts \cite{Undseth2023}. In this work, PIRS was observed through residual phase pick-up corresponding to a frequency shift on the order of \SI{10}{}-\SI{100}{\kilo\hertz}. Last, the measured PIRS phase pickup is consistent with a \textit{negative} Larmor frequency shift opposite to the frequency shift seen when raising the device temperature. This suggests that although the device temperature clearly plays a meaningful role in how PIRS manifests, the dissipated heat of the microwave burst may not be the only PIRS mechanism in this device. It has been proposed that thermally-sensitive environmental fluctuators can significantly contribute to the temperature dependence of the qubit frequencies \cite{Choi2024}. Our observations underscore the difficulty of predicting and compensating multi-qubit crosstalk with resonant control.

\section{\label{sec:-15mT}Baseband Operation}

\begin{figure*}
    \centering
    \includegraphics[width=\textwidth]{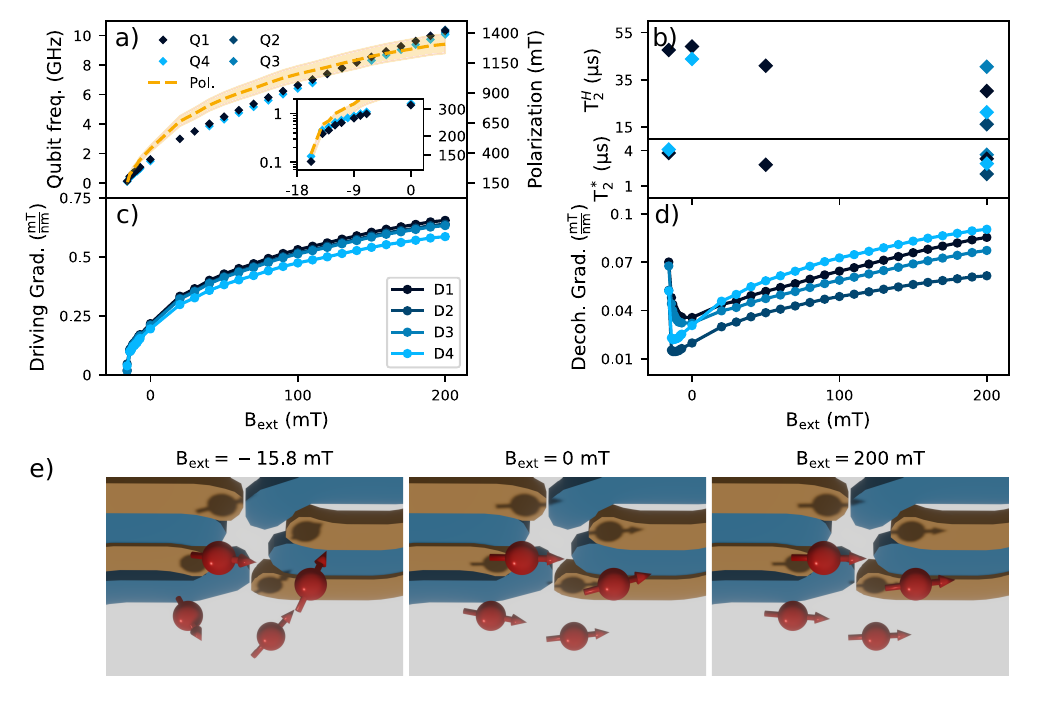}
    \caption{\textbf{Magnetic Field Dependence of Device and Qubit Properties} 
    a) Dependence of the qubit frequencies on the external magnetic field. Adiabatic inversion pulses are used for external fields above $B_{\rm ext}=\SI{-15}{\milli\tesla}$ (qubit frequencies above \SI{400}{\mega\hertz}) while Ramsey experiments with hopping control is used for lower fields. The dashed line shows the predicted homogeneous polarization of the micromagnet as extracted by fitting the measured qubit frequencies to a simplified micromagnet model (see Supplementary Information Section~II). The shaded area provides a generous estimate of uncertainty by displacing the micromagnet model $\pm$\SI{15}{\nano\meter} out-of-plane and repeating the fits. The inset zooms in onto the behavior below $B_{\rm ext}=\SI{0}{\milli\tesla}$ with the qubit frequencies and polarization plotted on a logarithmic scale. 
    b) (lower) $T_2^*$ integrated for 30 minutes at various external field settings. Ramsey decay times are relatively unchanged by the magnitude of the external field and, consequently, the decoherence gradient. Symbols as in panel a). 
    b) (upper) $T_2^\mathrm{H}$ integrated for 30 minutes at various external field settings showing an inverse relation to external field and decoherence gradient. From the extracted micromagnet polarization in panel a) we extract 
    c) the driving gradient along the y-axis and 
    d) the decoherence gradient magnitude at the center of the plunger gates. The driving gradient decreases monotonically while the decoherence gradient shows a sweet spot close to $B_\mathrm{ext}=\SI{0}{\milli\tesla}$.
    e) Renders of the plunger and barrier gates (yellow, blue) from below with the four electron spins (red, from left-to-right: Q3, Q2, Q4, Q1) for three different external field settings. The arrows indicate the quantization axes at the dot location which, approximating an isotropic $g$-factor of 2 in the silicon quantum well, is effectively assumed to be colinear with the total magnetic field vector. At $B_\mathrm{ext}=\SI{200}{\milli\tesla}$ the quantization axes are almost perfectly aligned with each other within a few degrees. At $B_\mathrm{ext}=\SI{0}{\milli\tesla}$, some deviation becomes evident. At $B_\mathrm{ext}=\SI{-15.8}{\milli\tesla}$ the differences in quantization axes are very pronounced. Based on the micromagnet simulation, the tips are estimated to be between about \SIrange[range-phrase = {\text{~and~}}]{30}{60}{\degree} depending on the exact locations of the quantum dot in the micromagnet stray field.}
    \label{Fig:xvsB_umag}
\end{figure*}

In light of the crosstalk challenges arising from microwave control, the operation of spin qubits using exclusively baseband pulses makes for a compelling alternative while retaining the Loss-DiVincenzo qubit encoding. Such control was recently demonstrated by hopping a hole spin qubit between germanium quantum dots with different quantization axes \cite{Wang2024} and bears resemblance to the strong-driving flopping mode regime proposed by \cite{Teske_2023}. Similarly, the 2x2 silicon array allows us to take advantage of quantization axis variation originating from the inhomogeneous magnetic field introduced by the micromagnet. However, even in the 2x2 array a large external field forces the spins' quantization axes to be nearly colinear. In order to accentuate the inhomogeneous magnetic field and limit Larmor frequencies for baseband control, the external field is reduced.

First, the micromagnet is magnetized in a \SI{0.8}{\tesla} external field. We populate the array with a single electron in dots 1 and 4, allowing us to keep PSB readout on the pair Q$_4$Q$_1$ while dots 2 and 3 remain empty. We track the qubit frequencies of Q$_1$ and Q$_4$ using adiabatic inversion pulses as the external field is reduced step-by-step as shown in Fig.~\ref{Fig:xvsB_umag}a). We observe the qubit frequencies dropping both as a result of the lower external field as well as the demagnetization of the micromagnet. The polarization of the micromagnet is inferred by fitting the measured frequencies to a simplified magnet model where the magnetization is assumed to be homogeneous with no crystalline domains or shape anisotropy coming into consideration (see Supplementary Information Section~II). Even so, this model provides a firm basis for understanding the qualitative qubit properties across a wide range of external field settings.

Even at zero external field, the remanence of the micromagnet forces the quantization axes to be reasonably colinear and provides a driving gradient large enough for conventional EDSR operation. Notably, the Hahn-echo coherence times of Q$_1$ and Q$_4$ nearly double from \SI{30.2}{\micro\second} and \SI{21.3}{\micro\second} to \SI{49.1}{\micro\second} and \SI{43.8}{\micro\second} respectively as shown in Fig.~\ref{Fig:xvsB_umag}b). Curiously, the $T_2^*$ coherence times do not change significantly with the magnetic field and decoherence gradient. This suggests the limiting low-frequency noise may be different than the noise inducing Hahn-echo decay. The amplitude with which these spins are driven with $f_\mathrm{Rabi}\approx$\SI{2}{\mega\hertz} also increases by about a factor of 3. These observations are compatible with the estimated change in decoherence and driving gradients at the dot locations: Fig.~\ref{Fig:xvsB_umag}c) shows that the decoherence gradient permitting charge noise to couple to the spins has approximately halved while Fig.~\ref{Fig:xvsB_umag}d) shows the driving gradient has been reduced by a factor of 3. As predicted by simulations, the \SI{100}{\mega\hertz} frequency separation between Q$_1$ and Q$_4$ is retained and becomes the main contribution to the decoherence gradient. This suggests that EDSR operation in the absence of an external solenoid, a mundane but material obstacle for the near-term scalability of semiconductor spin qubits, is a realistic possibility.

Decreasing the external field further, we reach a regime where the local magnetic field in the dot region is increasingly inhomogeneous owing to the competition between the remaining micromagnet polarization and the opposed external field. The spin quantisation axes at the quantum dots will no longer align but instead point in different directions. This is most clearly illustrated by rendering the quantization axes at each dot location as shown in Fig.~\ref{Fig:xvsB_umag}e). Despite the extreme inhomogeneity and low effective polarization of the micromagnet, we observe no pronounced instability of the spin frequencies. On the contrary, the measured Ramsey and Hahn-echo decay times universally improve upon the coherence measured at $B_\mathrm{ext}=\SI{200}{\milli\tesla}$ with EDSR control (Fig.~\ref{Fig:xvsB_umag}b)). However, at a nominal external field setting of $B_\mathrm{ext}=\SI{-15.8}{\milli\tesla}$ the driving gradient is too small for resonant control. Fortunately, the large quantization axis tips enable coherent control by hopping the electrons from dots D$_1$ and D$_4$ to the vacant dots D$_2$ and D$_3$. Furthermore, the Q$_1$ and Q$_4$ Larmor frequencies of \SI{131}{\mega\hertz} and \SI{103}{\mega\hertz} respectively are practical for AWG control.

\begin{figure*}[htbp]
    \centering
    \includegraphics[width=0.91\textwidth]{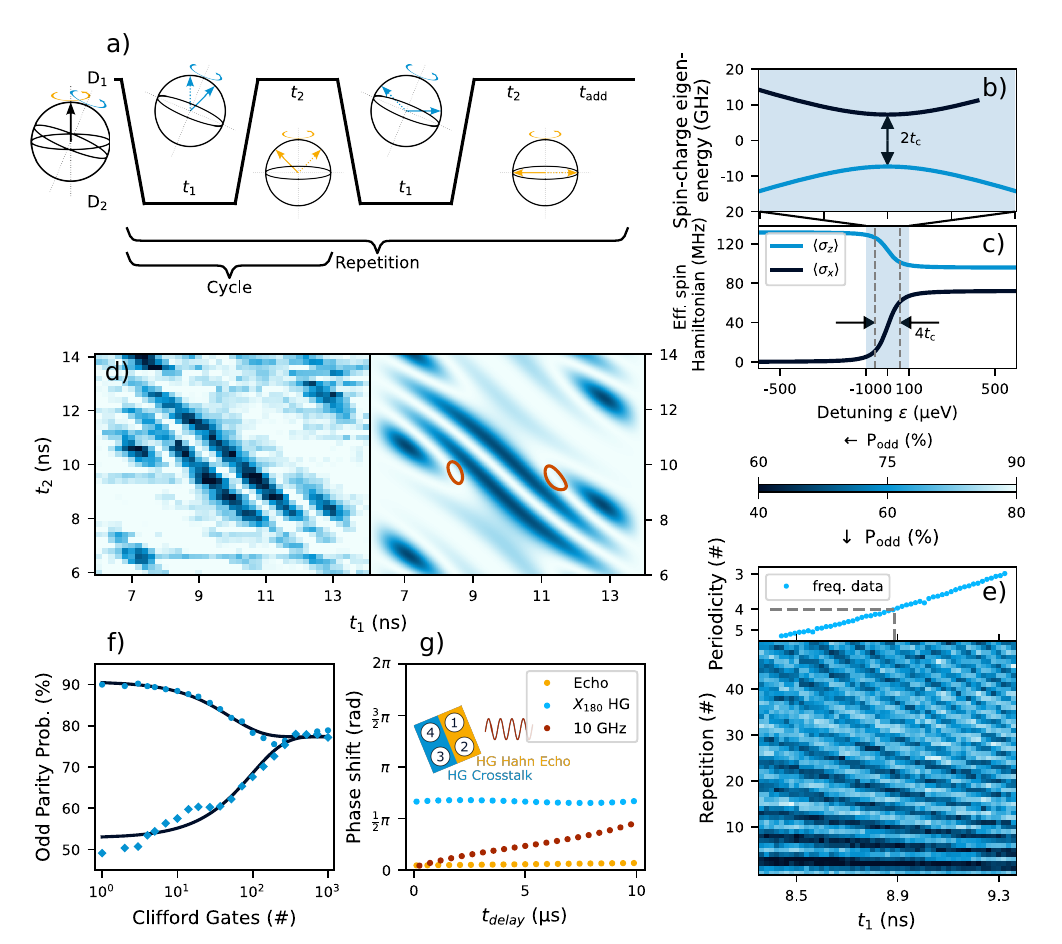}
    \caption{\textbf{Implementation and Characterization of a Hopping Gate} a) Hopping sequence and sketched evolution of the spin state vector at each step. The dashed (solid) arrow indicates the initial (final) vector at each step while yellow (blue) correspond to the original (secondary) quantum dot. The sequence results in a 90 degree rotation around the x-axis in D$_1$.
    b) Spin-charge energy level diagram for the D$_{1,2}$ double-dot system illustrating the charge avoided-crossing. The spin-splitting is too small to resolve compared to the much larger energy scale of the tunnel coupling.
    c) Components of the effective spin qubit Hamiltonian $H'$ with respect to the quantization axis of D$_1$ in the subspace of the charge ground state given by $\langle \sigma_i\rangle = \mathrm{Tr}(\sigma_i H')$ for the D$_{1,2}$ double-dot system (see Supplementary Information Section~V).
    d) Experimental (left) and fitted (right) odd-parity probability using two cycles and four repetitions [refer to (a)] while sweeping $t_1$ and $t_2$ for the D$_{1,2}$ double-dot system with $t_\mathrm{add}$ fixed to \SI{3.6}{\nano\second}. The quantization axis tip is fit to \SI{37.3(2)}{\deg} with a unitary model (see Methods for details). A precise combination of $t_1$, $t_2$ and $t_{\rm add}$ will create an accurate X$_{90}$. The red regions highlight the $t_1$ and $t_2$ where each one of the four repetitions achieves a high fidelity X$_{90}$ gate. Additional data, fits and fidelity contours for one, two, and four repetitions of qubits Q$_1$ and Q$_4$ are shown in Extended Data Fig.~\ref{Ext_Fig:HG_Fitting}.
    e) Fine calibration of $t_1$ by repeating the roughly calibrated gate many times with minor adjustments to one of the timing parameters. We extract the periodicity of the pattern and select the $t_1$ that matches a periodicity of 4 for the benchmarked gate. The difference in $t_1$ with respect to the experimental data in c) is due to adjustments in $t_{\rm add}$.
    f) Randomized benchmarking using both initial basis states. The Clifford set is compiled using only X$_{90}$ hopping gates and physical Z-rotations (see Methods). 
    The fitted Clifford gate fidelities are $F_\mathrm{Clif}^\mathrm{hop,odd}= \SI{99.01(11)}{\percent}$ and $F_\mathrm{Clif}^\mathrm{hop,even}= \SI{99.49(7)}{\percent}$ for the odd and even parity inputs respectively.
    g) PIRS-like measurements showing the phase accumulation of a hopping gate and a \SI{10}{\giga\hertz} burst when included in an echo sequence as depicted in Fig \ref{Fig:EDSR} b). The hopping gate is implemented as an X$_{180}$ gate on Q$_4$. The microwave burst carries the same energy as an average X$_{90}$ gate in this device during conventional EDSR operation.
    }
    \label{Fig:DSG_PIRS_RB}
\end{figure*}

Fig.~\ref{Fig:DSG_PIRS_RB}a) illustrates the pulse sequence required for a hopping gate between dots  D$_1$ and  D$_2$. During each ramp, the charge transfer between dots should be adiabatic (see Fig.~\ref{Fig:DSG_PIRS_RB}b)), hence a sufficiently large tunnel coupling $t_c$ should be maintained between dots. A fast detuning ramp $v=\epsilon/t_\mathrm{ramp}$ is desirable to minimize dephasing while the spin is most strongly hybridized with the charge at $\epsilon=\SI{0}{\micro\eV}$. Within the subspace of the adiabatic charge transition, the spin state transfer should be diabatic (see Fig.~\ref{Fig:DSG_PIRS_RB}c)) which is ensured by traversing the Hamiltonian ``step'' sufficiently fast. Provided these conditions are met (see Supplementary Information Section~V), a universal gate set may be calibrated with the appropriate choice of parameters $t_1$, $t_2$, and $t_\mathrm{add}$ \cite{Wang2024}. One repetition of two cycles as drawn is sufficient to calibrate an X$_{90}$ provided the quantization axis tip between dots lies between \SI{22.5}{\deg} and \SI{157.5}{\deg}. For quantization axis tips between \SI{45}{\deg} and \SI{135}{\deg}, one repetition of a single cycle suffices. The additional idling time parameter $t_{add}$ between subsequent repetitions ensures the axis of rotation is consistent when concatenating multiple X$_{90}$ gates.

To quantify the quantization axis tip experimentally, we apply multiple repetitions of spin hopping for varying times $t_1$ and $t_2$ and measure the resulting spin fraction. The measured pattern sensitively depends on the tip angle and may be fit numerically to the expected unitary evolution of the spin (see Methods). An example of a 2-cycle-4-repetition-protocol measurement and fit is shown in Fig~\ref{Fig:DSG_PIRS_RB}d) and yields a tip angle of $\theta_{1,2} = \SI{37.3(2)}{\deg}$. Similarly, we measure $\theta_{4,3} =\SI{47.5(2)}{\deg}$ (see Extended Data Fig.~\ref{Ext_Fig:HG_Fitting}). In the latter case, we observe that pulsing on P3 causes severe degradation to the readout signal (as was also the case for probing two-qubit interactions in Fig~\ref{Fig:Dev_Chev_DCZ}). We therefore opt to only benchmark the Q$_1$ hopping gate.

To coarsely calibrate the $t_1$, $t_2$ and $t_{\rm add}$ idling times for a high-fidelity single-qubit gate, we select parameters where the X$_{90}$ fidelity is theoretically maximized as shown in Fig.~\ref{Fig:DSG_PIRS_RB}d). We then fine tune the parameters by running a sequence of X$_{90}$ gates and select the precession times that produce a periodicity of 4 as exemplified for $t_1$ in Fig.~\ref{Fig:DSG_PIRS_RB}e). Consequently, we observe discretized Rabi oscillations of around \SI{5}{\mega\hertz}. By padding idling times with $2\pi$ rotations, we intentionally do not operate the gates as fast as is theoretically possible in order to minimize artifacts that occur at the limit of the AWG time resolution. Z-rotations are implemented with a physical wait time calibrated by a Ramsey measurement of the spin Larmor frequency. Constructing a Clifford gate set comprised of X$_{90}$ hopping gates and physical Z-rotations, single-qubit randomized benchmarking of Q$_1$ yields an average Clifford gate fidelity of about $F_\mathrm{Clif,odd}^\mathrm{hop} = \SI{99.01(11)}{\percent}$ and $F_\mathrm{Clif,even}^\mathrm{hop} = \SI{99.49(7)}{\percent}$ as shown in Fig.~\ref{Fig:DSG_PIRS_RB}f). By attributing all errors to the X$_{90}$ hopping gate we can provide a lower bound on its fidelity $F_\mathrm{X_{90}}^\mathrm{hop} = 99.50(6)\%$  (see Methods). This is directly comparable to the physical resonant gate fidelities achieved with EDSR control at high field. We estimate that the hopping gate fidelity is limited predominantly by the limited tunnel coupling tunability in this device leading to non-adiabatic charge shuttling (see Supplementary Information Section~V for further discussion).

We may also compare the performance of the EDSR and hopping mechanisms with regards to crosstalk (Fig.~\ref{Fig:DSG_PIRS_RB}g)). We employ the same PIRS methodology as in Fig.~\ref{Fig:EDSR} and perform a Hahn echo sequence on qubit Q$_1$ using hopping gates. As Q$_4$ is used for parity readout, we calibrate an X$_{180}$ hopping gate on this qubit and embed it within the echo sequence (in place of the off-resonant burst) to measure the resulting phase pickup from the physical shuttling. In contrast to the EDSR-based gates, the hopping mechanism imparts a constant phase pick-up regardless of its temporal position in the echo sequence. We attribute this shift to the change in electrostatic conditions caused by both the gate voltage pulses and movement of the Q$_4$ electron in the quantum well which shifts the position of Q$_1$ in the magnetic field gradient. The qubit pair is maintained with very little residual exchange, hence negligible conditional phase is acquired during the sequence.

We also introduce a \SI{10}{\giga\hertz} burst commensurate with an X$_{90}$ gate in the standard EDSR regime. We observe a transient phase pickup analogous to the experiments executed at high-field. Notably, the nonlinear ``masking'' due to the resonant echo pulse is not observed here as the decoupling is performed with a hopping gate. The total transient phase pickup due to PIRS is reduced compared to Fig.~\ref{Fig:EDSR}. This is likely a consequence of the low magnetic field environment, as either a $g$-factor change or electron displacement in the gradient will result in a smaller shift in qubit Larmor frequency. In any case, the shuttling gate avoids the transient effect of the microwave burst, and the magnitude of the effect would decrease as the distance between local operations is increased, making a low-field shuttling implementation of single-qubit gates an appealing approach to reduce crosstalk and improve multi-qubit control.

\section{\label{sec:Architect} Architectural Proposal}

The design of on-chip magnets for EDSR control of extensible 2D spin arrays presents a design challenge: engineering controllability and spectral addressability with a small pitch requires nanoscale magnetic variation. With magnets or current-carrying wires that are an order-of-magnitude larger than the quantum dots themselves, this is difficult to achieve. Proposals making use of nanomagnets, with features similar in size to a $\approx\SI{100}{\nano\meter}$ dot pitch, are a promising solution \cite{Aldeghi2023, Aldeghi2024, forster2015, Tadokoro2021, bersano2024, Legrand_2023}. 

Inspired by these ideas, nanomagnets may also be used for hopping control. Compared to the microscopic $g$-tensor variations in germanium quantum dot arrays and similar platforms, local magnetic field variations could be engineered more deterministically. It is therefore worthwhile to revisit the design requirements for on-chip nanomagnets with hopping control in mind. Two specifications stand out compared to EDSR. First, qubit addressability with hopping gates is more localized and does not rely on the spectral separation of qubit frequencies. Second, the transverse gradient required for EDSR control effectively mandates a total magnetic field magnitude on the order of tens of \unit{\milli\tesla}, whereas tips in the quantization axis for hopping control (i.e. inhomogeneity of the magnetic field vector) are possible at more modest magnetic fields as illustrated by this work.

\begin{figure*}
    \centering
    \includegraphics[width=\textwidth]{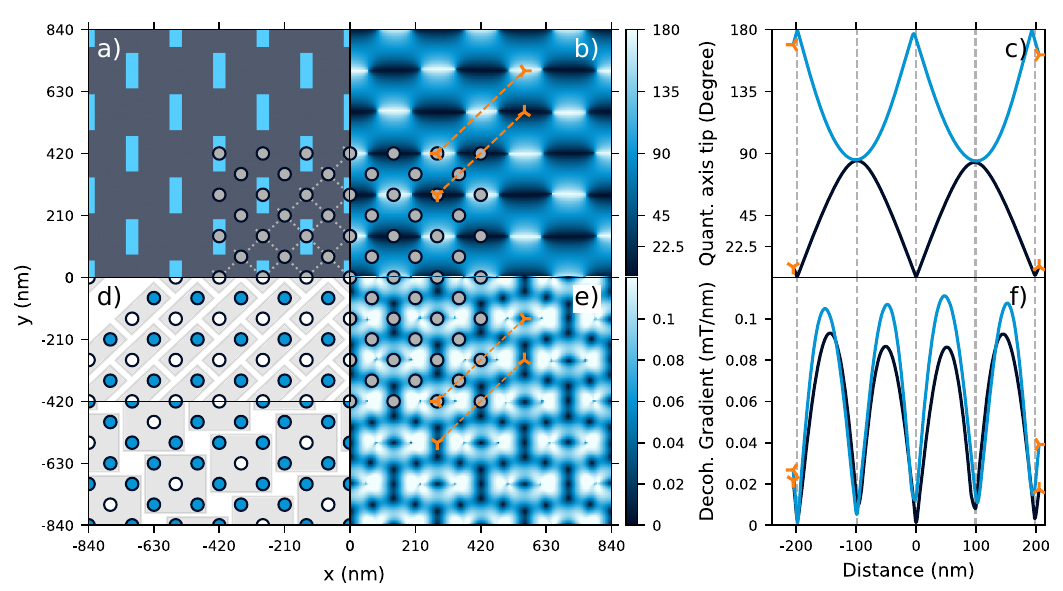}
    \caption{ \textbf{Nanomagnet Based Tileable Architecture Utilizing Hopping Gates} 
    a) Zoomed in section of the proposed nanomagnet pattern with the magnetic material shown in bright blue. Each iron nanomagnet is $\SI{40}{\nano\meter} \times \SI{120}{\nano\meter}$ in area and \SI{50}{\nano\meter} in out-of-plane height. In the lower right corner we show the envisioned quantum dot positions and their connectivity. These patterns extend across the contiguous panels a), b), d), e) and pertain to the data contained within them.
    b) The relative change in quantization axis, assuming an isotropic $g$-factor, with respect to the central dot of the panel. The quantum well is located \SI{100}{\nano\meter} below the bottom surface of the nanomagnet layer. The magnetic field is computed at zero external magnetic field after having allowed the nanomagnet magnetization to relax according to their shape anisotropy and material properties. 
    c) Line cuts corresponding to the orange markings in panel b) illustrating the relative change in quantization axis tips along two potential hopping axes. The observed asymmetries are a consequence of the finite mesh size used for simulation and interpolation artifacts. The grey dashed lines indicate the quantum dot positions. 
    d) Versatile filling of the quantum dot array: the bright blue filled dots indicate a quantum dot with an electron, the white markers indicate an empty quantum dot. The upper section shows a 50\% filling, the lower section a denser filling with 80\% occupancy. The gray shaded areas indicate the repeated unit cell. 
    e) The calculated decoherence gradient resulting from the nanomagnet stray field as described in b). 
    f) Line cuts corresponding to the orange markings in panel e) illustrating the decoherence gradient along two potential hopping axes. Minor artifacts from the finite mesh size and field interpolation are visible.
    }
    \label{Fig:Architecture}
\end{figure*}

Exploiting these features of hopping control, we propose a periodic nanomagnet design as depicted in Fig.~\ref{Fig:Architecture}a). After applying an external magnetic field along the y-axis, the shape anisotropy imposed by the $\SI{40}{\nano\meter} \times \SI{120}{\nano\meter} \times \SI{50}{\nano\meter}$ iron magnet geometry causes the magnetization to persist even at ambient conditions \cite{Aldeghi2023}. The relaxed magnetization, simulated by the OOMMF software package \cite{OOMMF} (see Supplementary Information Section~III), generates an inhomogeneous magnetic field that enables hopping conditions in a quantum well situated about \SI{100}{\nano\meter} below the magnetic layer. Owing to the inhomogeneous magnetic vector field, a predictable pattern of quantization axis tips (relative to a fixed reference point) of around \SI{90}{\deg} occurs with a periodicity of \SI{200}{\nano\meter} as shown in Fig.~\ref{Fig:Architecture}b). Fig.~\ref{Fig:Architecture}c) shows this smoothly-varying tip along two shuttling axes in which hopping gates may be implemented between adjacent sites with a pitch of about \SI{100}{\nano\meter}.

A sparse 2D array of silicon quantum dots may therefore be populated with flexible occupancy to realize different spin connectivity within the magnetic landscape. Fig.~\ref{Fig:Architecture}d) (upper) illustrates an instance where many dots may be left vacant to facilitate not only hopping gates but also shuttling to peripheral sensing regions or to implement two-qubit gates between distant spins. Fig.~\ref{Fig:Architecture}d) (lower) depicts a denser array where all qubits still have access to single-qubit logic via hopping. As any quantization axis tip between \SI{45}{\deg} and \SI{135}{\deg} allows Hadamard gates to be performed with a single shuttling cycle, this architecture is robust to some variation in both the placement/shape of the magnets and formation of the dots. As the nanomagnets can be fabricated over a large area, edge effects are relegated far beyond the extent of any finite quantum dot array. Of course, such a periodic array would also have to grapple with gate patterning and fanout as any other dense 2D array of dots would, but the issue of qubit controllability would not compound the difficulties. Provided such a fanout can accommodate the sparse magnetic layer, the design space for potential architectures becomes very large.

The nanomagnet array presented here confers two additional benefits beyond enabling hopping control that can also be leveraged in other potential designs. First, the qubit frequencies are distributed in a range below \SI{400}{\mega\hertz} without any applied external field. This removes the need for a solenoid in the cryogenic setup provided the magnets can be polarized at room temperature. Second, a periodic pattern of decoherence sweet spots also emerges to form a natural template upon which to place idling spins as seen in Fig.~\ref{Fig:Architecture}e). The proposed shuttling channels plotted in Fig.~\ref{Fig:Architecture}f) show that even the maximum decoherence gradient encountered in the quantum well is not substantially larger than typical EDSR addressability gradients that range from \SIrange{0.05}{0.1}{\milli\tesla/\nano\meter}. We note that different combinations of ferromagnetic materials and nanomagnet geometries may be used to achieve similar outcomes.

The hopping approach is not without drawbacks. High fidelity hopping control requires voltage pulses with very precise timing on the order of tens of picoseconds and mandates working with physical phase gates as opposed to virtualized phase tracking in a rotating frame. In the case of silicon, the microscopic valley degree of freedom may interfere with charge shuttling in cases where the valley phase difference between dots modifies the effective tunnel coupling\cite{Ginzel_2020, zhao2019coherentelectrontransportsilicon, Buonacorsi_2020, Teske_2023}. Also, the nanomagnets compete for space with the fanout of the gate electrodes that define the dots. Beyond engineering for single-qubit control, considering Zeeman energy differences for high-fidelity two-qubit gates \cite{Rimbach-Russ_2023} and Pauli spin blockade readout \cite{Takeda_2024} would be necessary to demonstrate a viable architecture. Engineering locations with zero magnetic field could also be useful for parking spins to alleviate the task of phase-tracking while idling qubits. Although we do not anticipate anything inhibiting these features with nanomagnet designs, we leave their details for future studies. 

\section{Conclusion}

We have used a 2x2 $^{28}$Si/SiGe quantum dot device to explore two ways in which future 2D arrays of single-electron spins in silicon may be controlled. Using microwave driving and barrier gate voltage pulses, we demonstrated universal control over a four-qubit system using conventional strategies from 1D arrays with only mild alterations. The performance was primarily limited by poor exchange control owing to the weak tunability of nearest-neighbour tunnel couplings. One potential remedy would be to modify the geometry of the barriers and invert the gate layers in fabrication to increase the lever arm controlling exchange. Another possibility would be the inclusion of a central gate to facilitate the decoupling of spins without requiring such large pulse amplitudes. Regardless, this demonstration of a 2D silicon spin array shows that established qubit control methods are sufficient for scaling beyond linear architectures. Longer bilinear arrays and perhaps trilinear devices ought to be compatible with established micromagnet designs. However, we also found that resonant single-qubit control causes crosstalk that complicates multi-qubit operation.

By opposing the polarization of the micromagnet with the external field to create a net vector field that varies over the scale of the array, we implement hopping gates that allow for baseband control of single spin qubits. Our initial characterization indicates several advantages of this operating regime. First, the baseband control pulses do not impart the transient phase crosstalk that was observed with resonant control. This may be attributed to the reduced power dissipation per logical operation which in turn reduces local heating effects \cite{Wang2024}. Second, the reduced magnetic field reduces the coupling strength to charge noise thus increasing the Hahn echo coherence substantially. Whereas the device was not optimized for hopping gate operation, we already demonstrate a hopping gate that is competitive with the EDSR control demonstrated in this device both in terms of speed and fidelity. We believe the same aforementioned improvements in tunnel coupling control would permit better fidelities in future implementations, and the gate speed would effectively double by increasing the tip of the quantization axis between dots to above \SI{45}{\deg}.

Finally, we explore how engineering on-chip magnets for hopping spins alleviates many of the challenges that impede scaling EDSR control. We put forward an illustrative example of a periodic nanomagnet pattern that creates a deterministic pattern of decoherence sweet spots and nearest-neighbour tip angles for implementing hopping spins in an arbitrarily large 2D array. These conditions can further enhance coherence times, increase the single-qubit gate speed, and do not require a superconducting solenoid to provide an external field. We believe further investigation into nanomagnet designs for hopping spin control would be very fruitful.

\section*{Acknowledgements}

We thank M. Aldeghi, C.-A. Wang, H. Tidjani, and I. Fernandez de Fuentes for very helpful discussions, S.L. de Snoo for software support, L. Tryputen and S. V. Amitonov for fabrication and design input as well as taking the presented SEM image, and O. Benningshof for technical support. Furthermore, we thank other members of the Vandersypen, Veldhorst, and Scappucci groups for their input and feedback. We acknowledge financial support from Intel, the Army Research Office (ARO) under grant number W911NF-17-1-0274 and W911NF2310110, and the Dutch the Ministry for Economic Affairs through the allowance for Topconsortia for Knowledge and Innovation (TKI). We also acknowledge support from the “Quantum Inspire–the Dutch Quantum Computer in the Cloud” project (Project No. NWA.1292.19.194) of the NWA research program “Research on Routes by Consortia (ORC),” which is funded by the Dutch Research Council (NWO).

\section*{Declarations}

The authors F.K. Unseld, B. Undseth, E. Raymenants, and L.M.K. Vandersypen are inventors on a patent application on nanomagnet designs for hopping spin control filed by Delft University of Technology under the application number NL2039255. The remaining authors declare that they have no competing interests.

\section*{Data Availability}
Data and scripts used in this publication are available in the Zenodo repository \cite{zenodo_dataset}.

\bibliography{sn-bibliography}

\FloatBarrier
\section*{Methods}
\label{sec:Methods}
\subsection*{Device Design and Fabrication}
The device layout used in this work are similar to the ones used in our previous work \cite{Unseld2023}. 
The heterostructure comprises a Si$_{0.69}$Ge$_{0.31}$ strained relaxed buffer (SRB), a \SI{7}{\nano\meter} tensile-strained $^{28}$Si quantum well (isotopic enrichment 800ppm), and a \SI{30}{\nano\meter} Si$_{0.69}$Ge$_{0.31}$ barrier passivated by an amorphous Si-rich layer \cite{DegliEsposti2024}. 

On top of the heterostructure we fabricate a multilayer gate stack isolated by a native SiOx layer and a \SI{10}{\nano\meter} thick layer of AlOx deposited using atomic layer deposition (ALD). The first metallic layer (3:\SI{17}{\nano\meter} Ti:Pd) connects to the ohmic contacts and defines screening gates. The screening gates define the area that will host the four quantum dots and two SETs and prevent electron accumulation or spurious dot formation in the fan out. Furthermore, the screening gates confining the 2x2 array are designed as coplanar waveguides to effectively deliver microwave signals for EDSR control.
The second layer (3:\SI{27}{\nano\meter} Ti:Pd) includes all plunger and accumulation gates, while the third layer (3:\SI{27}{\nano\meter} Ti:Pd) contains all barrier gates. The top layer of the gate stack (3:\SI{200}{\nano\meter} Ti:Co) contains only the micromagnet. Each layer is electrically isolated from the ones before by \SI{5}{\nano\meter} AlO$_{\rm X}$ grown with ALD.

The SETs are measured using RF reflectometry implemented using the split-gate approach \cite{Liu2021}. The size of the accumulation gates was substantially increased compared to \cite{Unseld2023} to increase the capacitance to the 2D electron gas (2DEG). The split accumulation gate could be depleted to create a high-resistance path from the 2DEG to the ohmic to prevent signal leakage.

The fan out of the barrier and plunger gates is shown in Extended Data Fig.~\ref{Ext_Fig:Set_Up}. The starred connections indicate the bond pads corresponding to B43, P3 and B23 which have uniquely circuitous traces to adapt to the distribution of fast-lines on the sample PCB. We consistently observe readout degradation when applying baseband pulses to these gates. As the cross-capacitance between these traces and adjacent bondpads (many of which connect to unvirtualized gates) is much larger in the fan out than at the scale of the quantum dot array, we suspect their suboptimal routing is correlated with the drop in readout quality.

\subsection*{Initialization, Control, and Readout}

A single-shot measurement includes initialization, manipulation and readout. Initialization and readout are performed using Pauli spin blockade and post-selection similar to \cite{Philips2022}. While PSB could be achieved in 4 qubit pairs in this device, we focus on pairs (4,1) and (2,3) as they offered superior performance. All shots begin with a round of parity-mode PSB measurement where parallel spin states are blockaded in the (1,1) charge state and antiparallel spin states are permitted to tunnel to the doubly-occupied singlet state S(2,0). A fast diabatic ramp is used to pulse from the readout point to the (1,1) charge state. When the post-selected initial state is S(2,0), the fast ramp yields a mostly-entangled state (see Extended Data Fig. \ref{Ext_Fig:QST}) with consistently higher single-qubit visibility than what is obtained with an adiabatic ramp to the (1,1) charge state. Prior to measurement, a wait-time of \SIrange{1}{5}{\micro\second} is included to permit both antiparallel spin states to decay to the S(2,0) state. The charge sensor signal is integrated for \SIrange{1}{20}{\micro\second}.

While a longer integration time improves readout quality somewhat, there appear to be other fundamental limits to the qubit visibilities we can achieve. One factor is the limited tunnel coupling tunability upon which the PSB mechanism depends very sensitively. Another factor is the opening of the S(2,0)-$\ket{\downarrow\downarrow}$ avoided crossing due to misaligned quantization axes which may impede a linear adiabatic ramp from the S(2,0) state to the $\ket{\downarrow\uparrow}$. This effect ought to be more pronounced during low-external-field operation when quantization axis tips are largest. Pulse-shaping and active feedback may be used to mitigate this, but these were not investigated in this work. Datasets are post-processed to shift the readout threshold to an optimal point.

Qubit manipulation includes resonant drive and diabatic spin shuttling for single-qubit gates. The Hamiltonian for the single-qubit gates are discussed in Supplementary Information Sections~I and V respectively. All microwave pulses are padded with \SI{5}{\nano\second} wait times to allow for a transient ring-down of the microwave source. Furthermore, the IQ amplitudes and phases are adjusted for each qubit at its center-frequency to suppress LO-leakage, unwanted sidebands and other spurious tones during IQ modulation. In all cases, a square pulse shape is used for the I and Q waveforms.

The exchange interaction, given by $H_\mathrm{exch} = 2\pi\hbar J\mathbf{S}\cdot\mathbf{S}$, is turned on and off via square barrier pulses for the DCZ oscillations shown in Fig.~\ref{Fig:Dev_Chev_DCZ}c). Due to the large Zeeman energy difference present between all qubit pairs, we take the two-qubit unitary evolution to be $U_\mathrm{CZ}(t) = \mathrm{diag}(1, \exp{(i\pi Jt)}, \exp{(i\pi Jt)}, 1)$.

The presented single- and two-qubit experiments were run at a wide variety of temperatures and fields. The chevron patterns in Fig. \ref{Fig:Dev_Chev_DCZ} b) and the randomized benchmarking in Fig. \ref{Fig:DSG_PIRS_RB} a) were run at \SI{250}{\milli\kelvin} and \SI{200}{\milli\tesla} to prevent heating effects. The decoupled CZ oscillations were measured at the same field but at base temperature (around \SI{20}{\milli\kelvin}). All external field sweeps and shuttling gate experiments were conducted at base temperature.

\subsection*{Randomized Benchmarking}

In this work, we benchmark single-qubit gates implemented via resonant control and hopping control. Here we discuss the implementation of the Clifford gate sets in greater detail to interpret the average gate fidelities that are extracted for both gate flavors. The reported error bars correspond to the standard deviations acquired from curve fitting.

When using resonant control, the X/Y compilation is used (see Table II of \cite{Xue2019}) whereby all Clifford gates are compiled into $\pm 90$ degree rotations about the X and Y axes of the Bloch sphere. A single microwave burst operation is calibrated: it has a constant amplitude, frequency, duration, and a post-burst padding of \SI{5}{\nano\second} to allow for the microwave source to ring down. The only difference between pulses is the relative phase with which the I and Q modulation channels begin. For a given sequence of Clifford operations, the error contributed by each physical gate should be nominally identical in the absence of any non-Markovian effects. As the X/Y compilation of the 24 Cliffords contains 52 primitive gates in total, the average resonant gate fidelity $F_\mathrm{avg}^\mathrm{res}$ from the set $\{X_{\pm 90}, Y_{\pm 90}\}$ is estimated to relate to the resonant Clifford gate fidelity $F_\mathrm{Clif}^\mathrm{res}$ as $F_\mathrm{avg}^\mathrm{res} = 1-(1-F_\mathrm{Clif}^\mathrm{res})\cdot\frac{24}{52}$. For Q$_1$ to Q$_4$ respectively, the Clifford gate fidelities are measured to be $F_\mathrm{Clif}^\mathrm{res} = \{\SI{99.28(5)}{\percent}, \SI{99.58(4)}{\percent}, \SI{98.7(2)}{\percent}, \SI{98.5(3)}{\percent}\}$. We use 20 random Clifford circuits per data point, and average the result of about 1200 post-selected single shot measurements per circuit. The corresponding average resonant gate fidelities are, respectively, $F_\mathrm{avg}^\mathrm{res} = \{\SI{99.67(2)}{\percent}, \SI{99.80(2)}{\percent}, \SI{99.39(8)}{\percent}, \SI{99.3(2)}{\percent}\}$.

When using hopping control, the X/Z compilation is used (see again Table II of \cite{Xue2019}) as the gate set is comprised of a unique X$_{90}$ hopping operation as well as Z rotations implemented with physical idling times. Negative rotation about the X axis is implemented by a Z$_{180}$ prior to a X$_{90}$ hopping gate. In contrast to the resonant gate compilation, the X and Z operations are of a fundamentally different nature and therefore contribute to the error rate in distinct ways. First, X$_{90}$ operations require more time than Z rotations and therefore pick up more error through dephasing. Second, X$_{90}$ operations require four precise intervals of Larmor precession whereas Z rotations require only one. This means that any uncorrelated jitter in the AWG ramps will cause more error in the X$_{90}$ operations. Third, as X$_{90}$ operations involve shuttling, errors may arise due to imperfect charge or spin transfer between dots. Therefore, X$_{90}$ gates should be more prone to error than any Z rotation.

Interleaved randomized benchmarking data was not taken to directly estimate individual gate fidelities, so we propose two ways of interpreting the fidelity of the individual physical operations. From randomized benchmarking, we obtain a Clifford gate fidelity $F_\mathrm{Clif}^\mathrm{hop}$. We use 250 random Clifford circuits per data point, and average the result of about 1500 (200 circuits) and 750 (50 circuits) post-selected single- shot measurements per circuit. The X/Z Clifford set is compiled from 89 physical gates in the set $\{X_{\pm 90}, Z_{\pm 90}, Z_{180}\}$, so one may estimate the average gate fidelity of this set as $F_\mathrm{avg}^\mathrm{hop} = 1-(1-F_\mathrm{Clif}^\mathrm{hop})\cdot\frac{24}{89}$. Alternatively, one may estimate a lower bound of the X$_{90}$ gate fidelity by attributing all of the error to the X$_{90}$ operations on the basis that all Z operations are effectively brief extensions to the preceding X$_{90}$ sequence. As each Clifford gate in the X/Z compilation contains exactly 2 X$_{90}$ gates, the X$_{90}$ fidelity may be bounded as $F_\mathrm{X_{90}}^\mathrm{hop} = 1-(1-F_\mathrm{Clif}^\mathrm{hop})\cdot\frac{1}{2}$. We run randomized benchmarking of the baseband gate set after initializing both in the odd and even parity states as shown in Fig.~\ref{Fig:DSG_PIRS_RB}(f). While both exponential trends decay to the same mixed state probability, the even parity branch contains a small oscillatory artifact that persists even with substantial averaging. While readout crosstalk or state leakage during the protocol may play a role, we are unsure of the true origin of this artifact. To be conservative in our evaluation of the X90 hopping gate quality, we extract the Clifford gate fidelity of $F_\mathrm{Clif,odd}^\mathrm{hop} = 99.01(11)\%$ for Q$_1$ from the odd parity decay. It follows that $F_\mathrm{avg,odd}^\mathrm{hop} = 99.73(3)\%$ and $F_\mathrm{X_{90},odd}^\mathrm{hop} = 99.50(6)\%$. This latter value is the estimated fidelity bound reported in the abstract. For the even parity branch we fit $F_\mathrm{Clif,even}^\mathrm{hop} = 99.49(7)\%$, $F_\mathrm{avg,even}^\mathrm{hop} = 99.86(2)\%$ and $F_\mathrm{X_{90},even}^\mathrm{hop} = 99.74(4)\%$.

We note that neither the X/Y or X/Z compilations used are maximally efficient at compiling Clifford operations from primitive operations. Therefore, the Clifford gate fidelities extracted from randomized benchmarking could themselves be improved by tailoring the gate set to the physical gates. The average gate fidelity estimates should therefore be the most reasonable quantitative benchmark with which to compare the gate performance.

\subsection*{Tip angle fitting}

There are multiple strategies to estimate the quantization axis tip $\theta$ between two dots. The visibility of oscillations observed after diabatically shuttling from one dot to another can be directly related to the tip in quantization axis \cite{van_Riggelen_Doelman_2024}. This provides a reasonable estimate if the maximum visibility of oscillations can be established and provides good contrast. Alternatively, the Larmor frequency as a function of detuning across the hopping double-dot pair can be measured and fit to Eq.~S25 \cite{van_Riggelen_Doelman_2024, Wang2024}. Both of these methods are made difficult in our case by not having EDSR control at the magnetic field condition at which hopping gates are possible.

Here, we take advantage of the fact that a number of repetitions $r$ of the shuttling sequence as drawn in Fig.~\ref{Fig:DSG_PIRS_RB}a) will result in a unique 2D ``fingerprint'' as a function of the timing parameters $t_1$ and $t_2$. Generally, the pattern will depend more sensitively on the tip $\theta$ as $r$ increases. Exemplary patterns for both dot pairs are shown in Extended Data Fig.~\ref{Ext_Fig:HG_Fitting}.

We may model the pattern using the two-level Hamiltonian given in Eq.~S27 and make the approximation that the change in the Hamiltonian as a function of interdot detuning is perfectly sharp given the nanosecond-scale detuning ramp times \cite{Wang2024}. The time evolution is therefore a product of unitary Larmor precessions. In the initial dot, this precession occurs at an angular frequency of $\omega_\mathrm{init}$ about a quantization axis $\hat{z} = (0,0,1)^T$. In the tipped dot, this precession occurs at an angular frequency of $\omega_\mathrm{tip}$ about a quantization axis $\hat{\theta} = (\sin\theta, 0, \cos\theta)^T$. Since the detuning ramp times are still finite, precession occurs during the ramps and can be accounted for by adjusting the nominal precession time such that $t_1' = t_1-t_{1,\mathrm{offset}}$ and $t_2' = t_2-t_{2,\mathrm{offset}}$. The unitary evolution for $r$ repetitions of the sequence shown in Fig.~\ref{Fig:DSG_PIRS_RB}a) is therefore given by:

\begin{multline}
    U(t_1', t_2', t_\mathrm{add}, \omega_\mathrm{init}, \omega_\mathrm{tip}, \theta) = \\
    \left(R_z(\omega_\mathrm{init}(t_2'+t_\mathrm{add}))R_\theta(\omega_\mathrm{tip} t_1')R_z(\omega_\mathrm{init} t_2')R_\theta(\omega_\mathrm{tip} t_1')\right)^r
\end{multline}

where $R_n(\alpha)=\exp(-i\alpha\hat{n}\cdot\vec{\sigma}/2)$ indicates a positive rotation about the unit vector $\hat{n}$ by an angle $\alpha$. The measured pattern is given by:

\begin{multline}
    \label{eq:pattern_fit}
    p_\mathrm{odd}(t_1', t_2', t_\mathrm{add}, \omega_\mathrm{init}, \omega_\mathrm{tip}, \theta, A, B) = \\A\mathrm{Tr}\left(OU\rho_\mathrm{odd}U^\dagger\right) + B,
\end{multline}

where $\rho_\mathrm{odd}$ is the initialized odd-parity state, $O=(\mathbb{1} - \sigma_z\otimes\sigma_z)/2$ is the odd-parity observable, and $A$ and $B$ are visibility and offset corrections due to SPAM errors.

We fit the experimental data to Eq.~\ref{eq:pattern_fit} with a pixel-by-pixel least-squares optimization over the parameters $t_{1,\mathrm{offset}}$, $t_{2,\mathrm{offset}}$, $\omega_\mathrm{init}$, $\omega_\mathrm{tip}$, $A$, $B$, and $\theta$. $t_\mathrm{add}$ is known exactly from the experiment definition. As the qubit frequencies, offset times, and visibilities are well-estimated from other experiments, they are optimized within narrow bounds. For $r=4$, the optimized value for $\theta$ is accurate within a degree. For fewer repetitions, the fit is less accurate, and we instead simulate using the $r=4$ fit directly.
\section*{Extended Data Figures}
\onecolumngrid

\setcounter{figure}{0}
\renewcommand{\figurename}{\textbf{Extended Data Fig}}  


\begin{figure*}[h!]
    \centering
    \includegraphics[width=11.5cm]{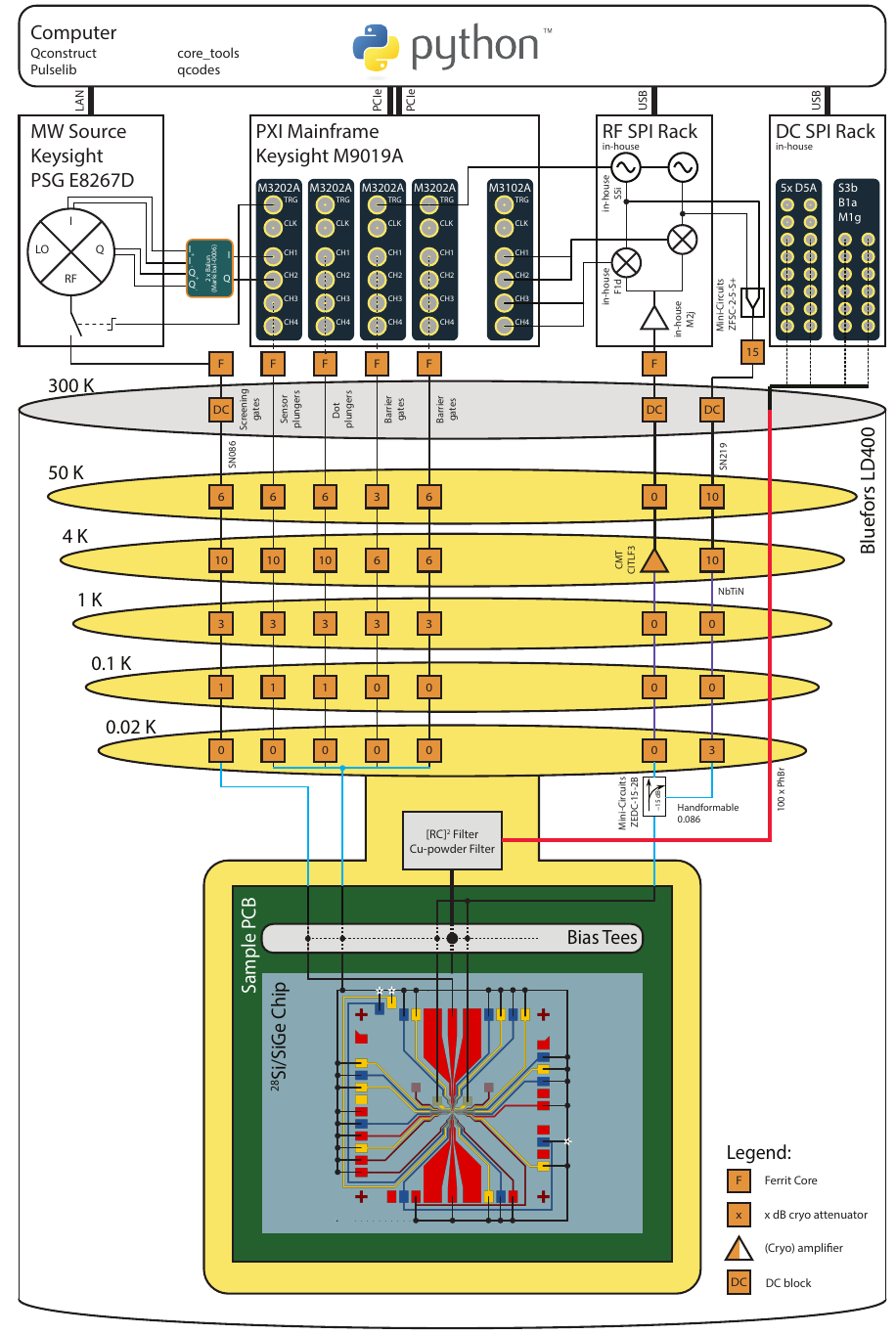}
    \caption{ \textbf{Schematic of the experimental setup} Arbitrary Waveform Generators (AWG, Keysight M3202A) in a PXI chassis are used to generate baseband pulses as well as I/Q input signals for the microwave (MW) vector source (Keysight PSG E8267D). Baseband signals are supplied to all plunger and barrier gates comprising the 2x2 array along with both sensing dot plunger gates. Lines routed to barrier gates are attenuated less to allow for larger voltage pulses on these gates. Wideband modulation with differential I/Q inputs is implemented using Balun's (Marki bal-0006). Ferrite cores are used to reduce low-frequency noise. Double DC blocks (Pasternack PE8210) are used on the microwave drive line and rf-reflectometry readout lines. DC voltages are supplied by battery-powered low-noise voltage sources and are filtered with low-pass RC filters and copper-powder filters before being combined on the sample PCB using bias tees with RC time constants of \SI{100}{\milli\second}. Two carrier frequencies are generated using homemade sources that are triggered by AWG marker channels and routed via bias tees to two readout tank circuits which are connected to accumulation gates adjacent to the two sensing dots using the split-gate method \cite{Liu2021}. The tank circuit consists of a series NbTiN inductor wirebonded on the PCB, the capacitance between the accumulation gate and the underlying 2DEG, as well as the resistive SET. The path from the 2DEG to the ohmic contact is depleted to minimize resistive leakage to ground. The reflected signal is separated from the incoming carrier with a directional coupler (Mini-Circuits ZEDC-15-2B), amplified at 4 Kelvin (CMT CITLF3) and room temperature before being demodulated and digitized (Keysight M3102A).}
    \label{Ext_Fig:Set_Up}
\end{figure*} 

\clearpage
\begin{figure}[h!]
    \centering
    \includegraphics[width=0.75\textwidth]{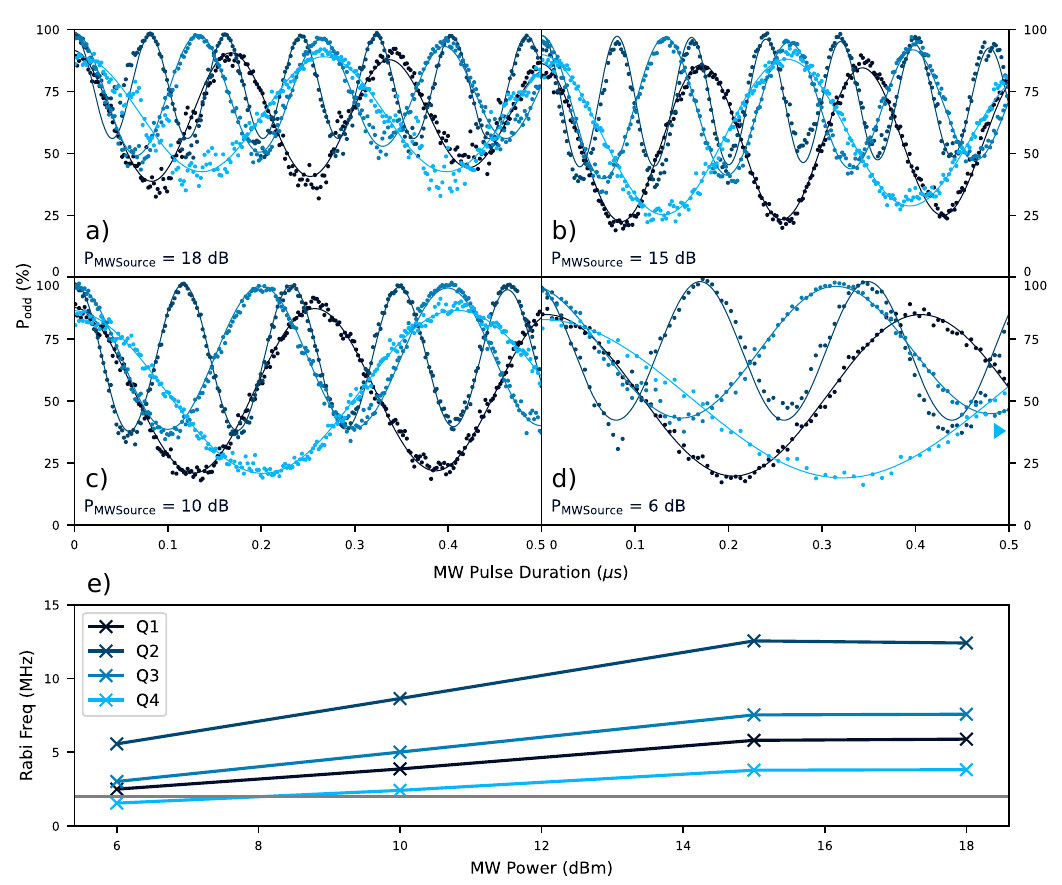}
    \caption{ \textbf{Rabi oscillations at various MW powers} 
    a)-d) Measured (data points) and fitted (solid lines) Rabi oscillations for all four qubits at various powers of the MW-source as annotated in the figures. We point out that the data of Q$_4$ in panel d) was cropped as indicated by the marker. The fit was performed with the full oscillation. 
    e) Rabi frequencies plotted against the MW-source powers. Above \SI{15}{\dBm} we observe a saturation on all qubits that is caused by the limitations of the microwave source. For the experiments in the main text we used \SI{6}{\dBm} and adjusted the IQ input amplitudes to achieve a Rabi frequency of around \SI{2}{\mega\hertz} (gray line) on all 4 qubits.}
    \label{Ext_Fig:EDSR_Driving}
\end{figure}
\clearpage
\begin{figure*}[h!]
    \centering
    \includegraphics[width=\textwidth]{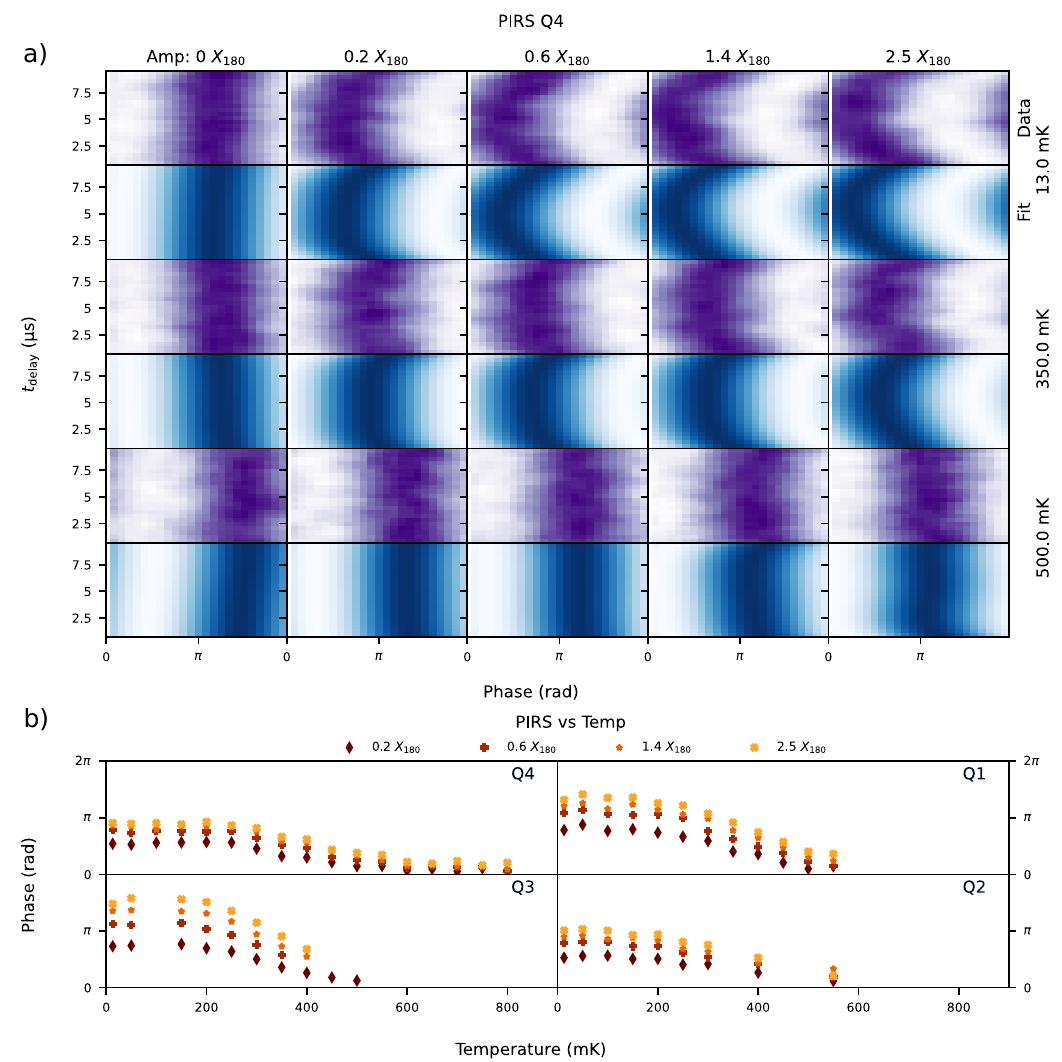}
    \caption{\textbf{PIRS during EDSR operation}
    a) Experimental data (purple) and respective fits (blue) for different off-resonant burst amplitudes at three exemplary mixing chamber temperatures for Qubit 4 (see Fig.~\ref{Fig:EDSR} for the pulse sequence schematic). As the individual traces at constant $t_\mathrm{delay}$ are both low in visibility due to the long echo pulse sequence used and subject to small signal drifts over the course of the long experiments, we fit the 2D datasets to Eq.~S24 using a quartic polynomial representation of $\theta_t(t_\mathrm{delay}) = \sum_{k=0}^4 c_k (t_\mathrm{delay}-t_0)^k$ to extract the smooth transient phase accumulation from the off-resonant burst as a function of $t_\mathrm{delay}$. In the left-most column, the experiment is a standard Hahn-echo sequence, and a constant phase offset is observed at the different temperature settings. We subtract this constant offset from the datasets plotted to the right which are collected at the same temperature in order to isolate the transient effect of the off-resonant burst. b) To quantify the relation between temperature and non-linear transient phase pick-up, we plot the maximum phase accumulation at as a function of mixing chamber temperature for all four qubits. All qubits exhibit the same behaviour whereby the maximum phase pickup is suppressed by operating at higher device temperatures. The positive phase accumulation is consistent with a negative qubit frequency shift given the definition of the rotating frame used in these experiments (see Supplementary Information Section~IV)}
    \label{Ext_Fig:EDSR_PIRS}
\end{figure*}

\clearpage
\begin{figure*}[h!]
    \centering
    \includegraphics[width=\textwidth]{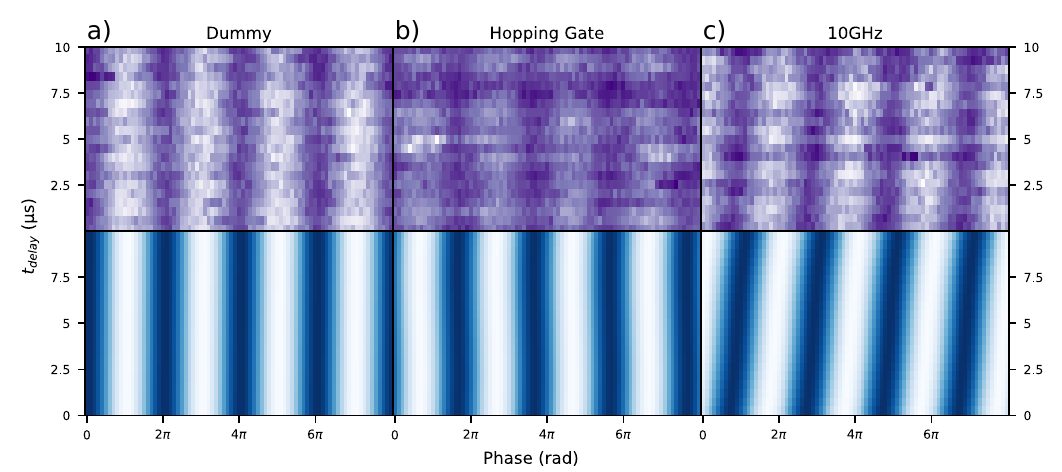}
    \caption{ \textbf{PIRS during baseband operation}
    a-c) Experimental data (purple) and respective fits (blue) using three variants of the PIRS experiment. a) The Hahn-echo pulse template is implemented on Qubit 1 using hopping gates. The same fitting procedure as described in Extended Data Fig.~\ref{Ext_Fig:EDSR_PIRS} is used. No off-resonant burst or pulse is applied. 
    b) An X$_{180}$ hopping gate is applied to Qubit 4 in place of the off-resonant burst. Qubit 1 experiences a non-transient phase pickup in addition to a \SI{180}{\deg} phase shift due to the change in measurement parity. The visibility is notably degraded compared to the other experiments due to the effect of baseband pulsing on gate P3. 
    c) A \SI{10}{\giga\hertz} burst is applied with an amplitude and duration energetically commensurate with 0.6X$_{180}$ during standard EDSR operation. A transient phase pickup is observed, but no ``saturation'' effect as in Extended Data Fig.~\ref{Ext_Fig:EDSR_PIRS} is seen. We believe this is due to the implementation of the decoupling pulse with a hopping gate rather than a resonant gate. The hopping gate itself does not impart a PIRS effect on the device.}\label{Ext_Fig:HG_PIRS_Fitting}
\end{figure*}
\clearpage
\begin{figure*}[h!]
    \centering\includegraphics[width=\textwidth]{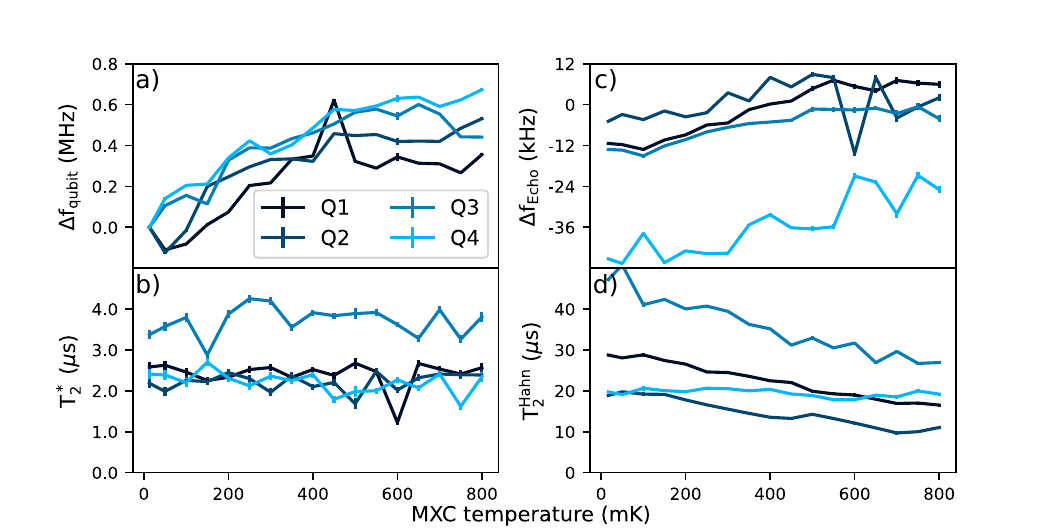}
    \caption{ \textbf{Temperature dependence of $T_2^*$, $T_2^\mathrm{H}$ and qubit frequencies} 
    a-b) Data extracted from Ramsey experiments (approx. \SI{30}{\minute}) conducted using EDSR control as a function of mixing chamber temperature. The measured oscillations can be fit to extract (a) the qubit frequency and (b) the Ramsey decay time $T_2^*$ per Supplementary Information Section~IV. Discounting spurious jumps in qubit frequency on the order of \SI{100}{\kilo\hertz} which we attribute to slow charge fluctuators, all qubit frequencies increase monotonically as a function of temperature within the range we are able to measure. The Ramsey decay times are notably unaffected by the device temperature for all four qubits. c-d) Hahn-echo experiments (approx. \SI{30}{\minute}) conducted using EDSR control as a function of mixing chamber temperature. The measured oscillations can be fit to extract (c) the difference in qubit frequency before and after the decoupling pulse and (d) the Hahn-echo decay time $T_2^\mathrm{H}$ per Supplementary Information Section~IV. The systematic difference in qubit frequencies is predominantly attributed to the PIRS induced by the decoupling X$_{180}$ pulse as it has a greater duration than the initial X$_{90}$ pulse. This systematic frequency difference appears to converge to zero as the mixing chamber temperature is increases, which is consistent with the observation that an increased device temperature mitigates the PIRS effect (see Extended Data Fig.~\ref{Ext_Fig:EDSR_PIRS}). Hahn-echo decay times for all qubits except Qubit 4 monotonically decrease as a function of mixing chamber temperature. Qubit 4's coherence remains constant with the changing temperature.}
    \label{Ext_Fig:T2vTemp}
\end{figure*}
\clearpage
\begin{figure*}[h!]
    \centering
    \includegraphics[width=\textwidth]{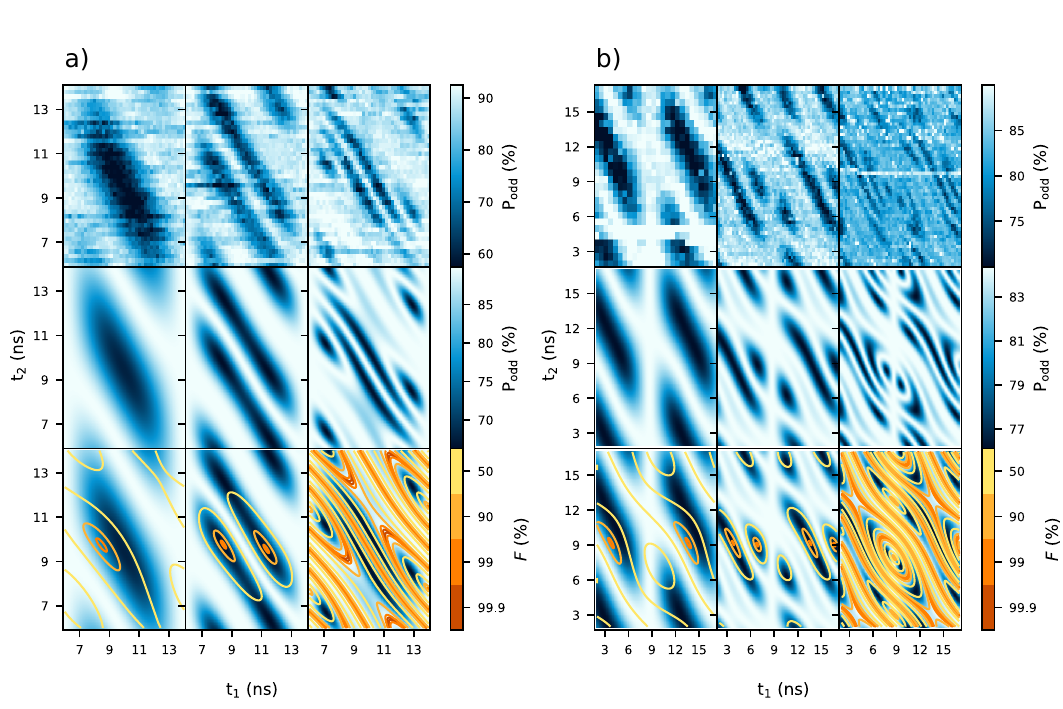}
    \caption{ \textbf{Hopping gate characterisation of D1D2 and D4D3} 
    Raw data (row 1), simulated data (row 2) and estimated gate fidelity (row 3) of one (column 1), two (column 2) and four (column 3) shuttle repetitions for the qubit in D1 (a) and the qubit in D4 (b). In both cases we used two shuttle cycles per repetition. To generate the simulated data sets we fit the four-repetition pattern and extract parameters such as the quantisation axis tip and timing offsets due to finite ramp times. Using the simulated data we calculate the gate fidelity as indicated by the contour plots: for one repetition, we calculate the X$_{90}$ fidelity; for two repetitions, we calculate the X$_{180}$ fidelity; and for four repetitions, we calculate the X$_{360}$ fidelity. The fidelity is calculated by assuming instantaneous shuttling between dots and considering only unitary spin state evolution. In the case of shuttling with Qubit 4, we observe the degradation of the readout signal due to the repeated pulsing on gate P3 (see Methods)}
    \label{Ext_Fig:HG_Fitting}
\end{figure*}
\clearpage
\begin{figure*}[h!]
    \centering
    \includegraphics[width=0.90\textwidth]{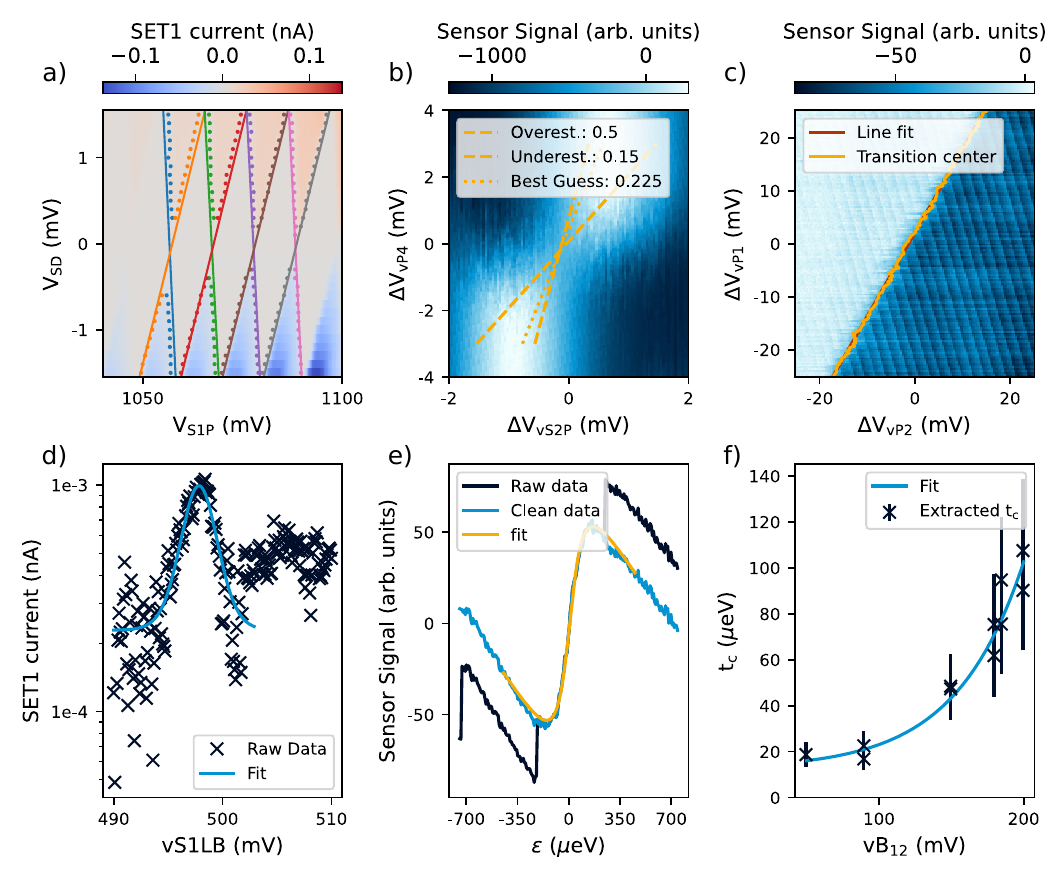}
    \caption{ \textbf{Lever arm, electron temperature and tunnel coupling measurements} 
    a) Coulomb diamonds of SET 1 showing the extracted slopes used to calculate the lever arm of the sensing dot plunger. 
    b) An exemplaray interdot transition of sensing dot 2 and quantum dot 4. The shift of the coulomb peak (bright white line) due to the added charge on quantum do D$_4$ is used for determining the lever arm ratio between the sensing dot plunger and the quantum dot plunger. Dashed and dotted lines show the manual over- and under-estimation as well as the best fit. The latter takes into account that both lever arms for vS1P and vS2P have to be in an agreement when propagated through the array.
    c) An exemplary interdot transition between quantum dots 1 and 2 used for determining the lever arm ratio between the two plunger gates. We fit the interdot transition line-by-line to extract a consistent center point followed by a linear fit through all center points to determine the slope. 
    d) Coulomb peak and fit to extract a bound on the electron temperature of about  $\mathrm{T}_{\rm e}<\SI{175}{\milli\kelvin}$. 
    e) Exemplary fit of an interdot transition to extract the tunnel coupling \cite{DiCarlo_2004}. The detuning lever arm as computed in panels a-c) as well as the electron temperature from d) are used in this fitting procedure. The sensing dot plunger is virtualized with respect to the detuning, and discrete steps in the measured signal due to finite AWG resolution are extracted from the measured transition prior to fitting.  
    e) Extracted tunnel couplings between dots 1 and 2 plotted against the respective barrier voltage showing the expected exponential dependency.}
\end{figure*}
\begin{table*}[htb]
    \centering   
    \caption{Lever arms in the 2x2 array. Virtualized sensor dot plungers (vS1P and vS2P) and virtualized loading barriers (vS1LB and vS2LB) are extracted directly via Coulomb diamonds. Virtualized quantum dot plungers (vP1 - vP4) are inferred by propagating the slope of interdot transitions through the array. The uncertainty of the virtual plunger gates is dominated by the uncertainty of propagating the sensor lever arm into the 2x2 array.}
    \label{tab:lever_arms}
    \begin{tabular}{c|c}
         Gate & Lever arm  \\\hline
vS1P & $0.1542\pm0.0017$\\\hline
vS1LB & $0.0155\pm0.0004$\\\hline
vS2P & $0.1537\pm0.0029$\\\hline
vS2LB & $0.0151\pm0.0002$\\\hline
vP1 & $0.0289\pm0.0083$\\\hline
vP2 & $0.0455\pm0.0131$\\\hline
vP3 & $0.0290\pm0.0084$\\\hline
vP4 & $0.0349\pm0.0101$\\\hline
    \end{tabular}
\end{table*}

\clearpage
\begin{figure}[h!]
    \centering
    \includegraphics[width=.90\textwidth]{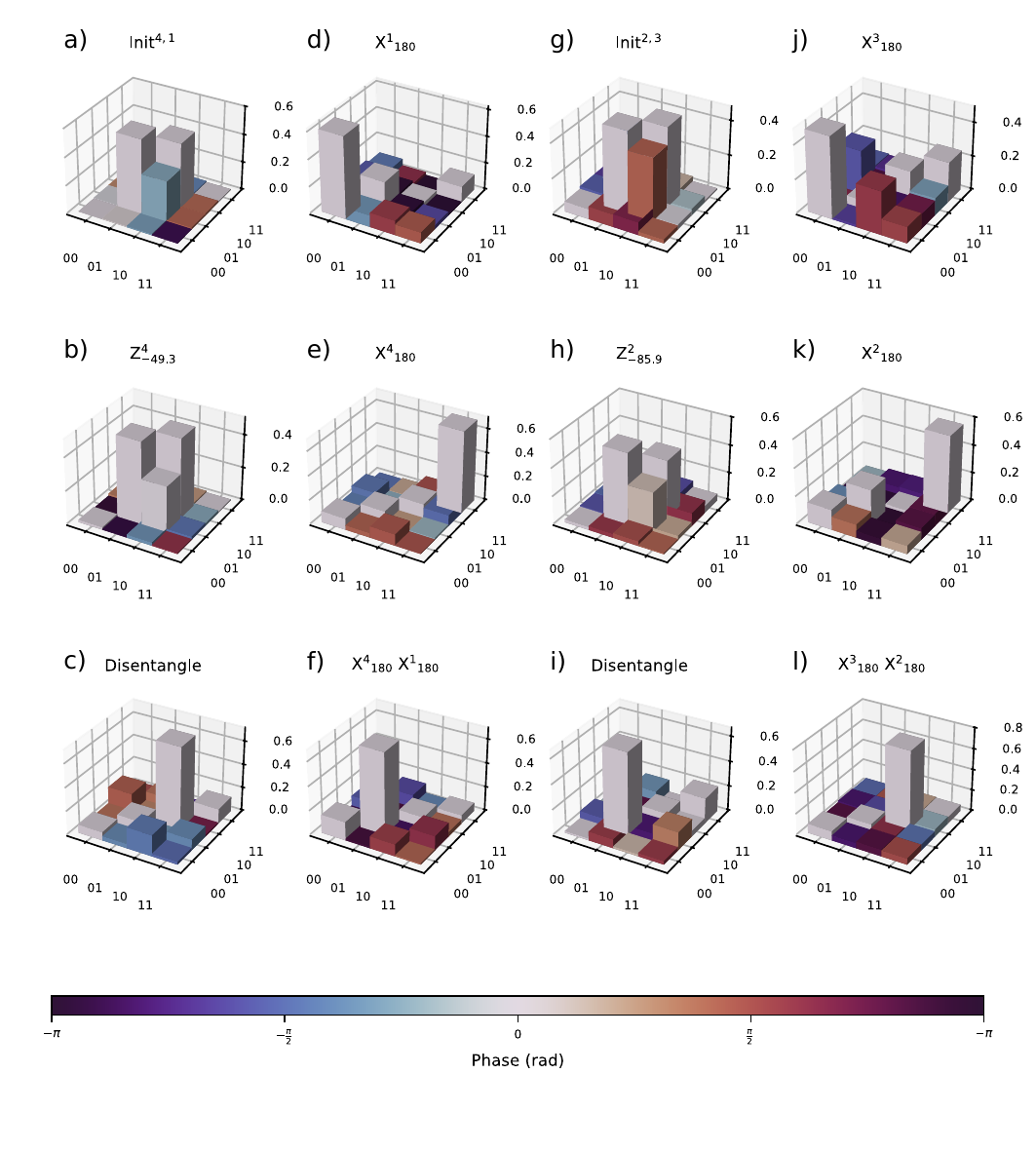}
    \caption{ \textbf{Quantum State Tomography of Qubit Pairs Q4Q1 (a-f) and Q2Q3 (g-l) obtained using EDSR and CZ operations} a)/g) Tomography of the initial state. As we transition quickly out of the (0,2)/(2,0) charge state to the (1,1) charge state, we initialise predominantly in an entangled state in the $\{\ket{01},\ket{10}\}$ subspace. This diabatic initialization offers more consistent state preparation compared to slow adiabatic ramping. In both cases we still observe a significant classical mixture of the two odd-parity states. For the single-qubit measurement protocols used in this work, a mixed initial state has no impact on the final result. However, such poor initialization has a large impact on most multi-qubit experiments. b)/h) The initialized state after applying a phase operation to produce a $\ket{T_0}$ state. c)/i) The initialized state after disentangling the qubits with a CZ and H gate. d)-f)/j)-l) The disentangled initial state after various X$_{180}$ gates to transform between the product states. A full two-qubit gate set is required to implement quantum state tomography on both Pauli spin blockade pairs illustrating full quantum logic on this device. The poor fidelity of the initial state as well as the poor quality factors of the two-qubit interactions limit the quality of further multi-qubit characterization on this particular device.}
    \label{Ext_Fig:QST}
\end{figure}

\clearpage
\begin{figure}[h!]
    \centering
    \includegraphics[width=0.9\textwidth]{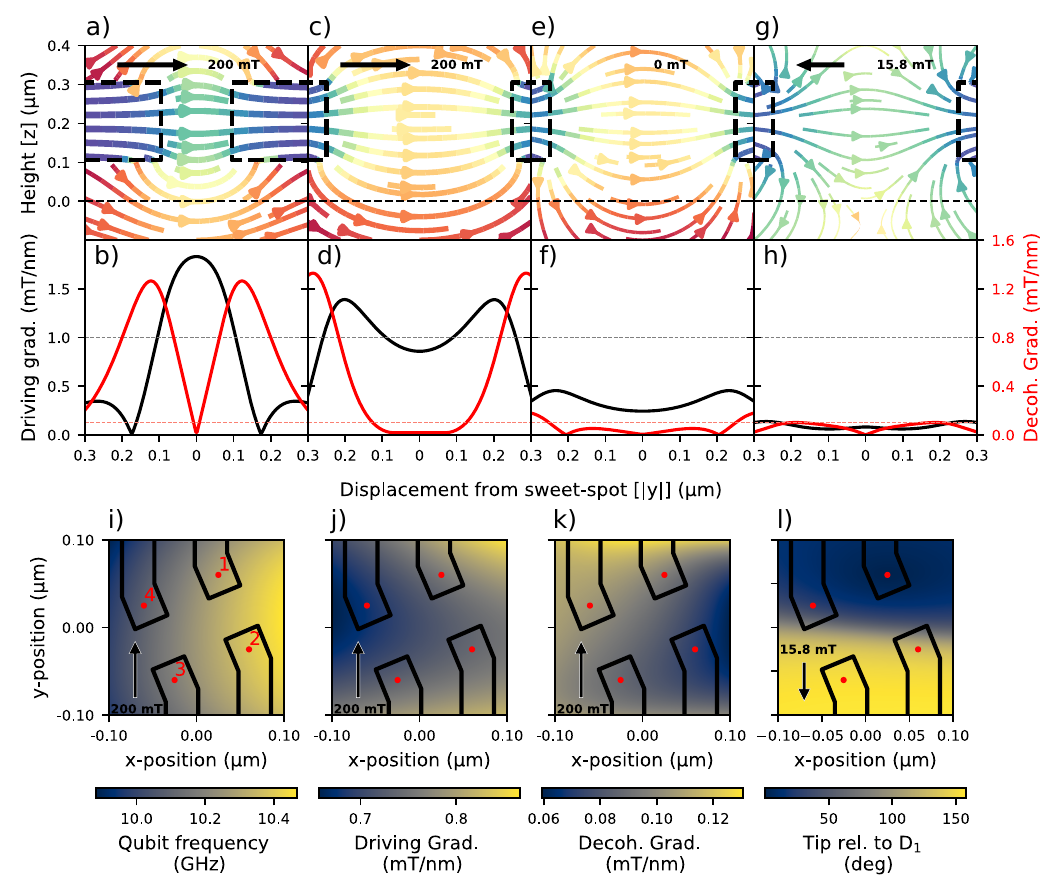}
    \caption{ \textbf{Micromagnet Simulations} a) Magnetic field lines forming in a micromagnet gap of \SI{190}{\nano\meter} as used in \cite{Philips2022} when an external field of \SI{200}{\milli\tesla} is applied as indicated. No addressability features along the x-axis are included in the magnet design, so the field lines only exist in the y-z plane. The dashed line indicates the level at which the 2DEG and quantum dots would be formed below the magnet. b) The simulated driving gradient (black) along the y-axis and estimated decoherence gradient (red) corresponding to the magnet design in a). The black dashed line indicates the \SI{1}{\milli\tesla/\nano\meter} minimum threshold for good EDSR control, and the red dashed line indicates the \SI{0.1}{\milli\tesla/\nano\meter} maximum threshold for good charge-noise resilience. Although a good driving gradient is present, the decoherence sweet spot is only wide enough to support a linear qubit array. c) Magnetic field lines forming in a micromagnet gap of \SI{500}{\nano\meter} similar to the design used in this work when an external field of \SI{200}{\milli\tesla} is applied as indicated. d) The simulated driving and decoherence gradients corresponding to the magnet design in c). Good EDSR is generally possible, and the decoherence sweet spot is wider than in a-b). e) Magnetic field lines forming in a micromagnet gap of \SI{500}{\nano\meter} when no external field is applied. f) The simulated driving and decoherence gradients corresponding to the magnet design in e). EDSR is only possible with a stronger driving field, but the sweet spot in which to form good qubits is substantially wider. g) Magnetic field lines forming in a micromagnet gap of \SI{500}{\nano\meter} when an external field of \SI{15.8}{\milli\tesla} is applied as indicated. h) The simulated driving and decoherence gradients corresponding to the magnet design in g). EDSR is effectively impossible due to the small gradient, but qubits may be controlled via shuttling. The decoherence sweet spot is still wide. i) Simulated qubit frequencies using the nominal micromagnet design used in this work and an external field of \SI{200}{\milli\tesla}. The polarization of the micromagnet used in simulation is inferred based on the measured qubit frequencies. The gate overlay indicates the approximate locations where Qubits 1-4 are formed in the magnetic field. j) The simulated driving gradient along the y-axis with an external field of \SI{200}{\milli\tesla}. k) The simulated decoherence gradient with an external field of \SI{200}{\milli\tesla}. l) The simulated tip in quantization axis relative to Dot 1 with an external field of \SI{15.8}{\milli\tesla} oriented opposite the micromagnet polarization illustrating the substantial tips formed between dots D$_1$D$_2$ and D$_3$D$_4$.}
    \label{Ext_Fig:micromagnet}
\end{figure}

\clearpage
\begin{figure}[h!]
    \centering
    \includegraphics[width=.9\textwidth]{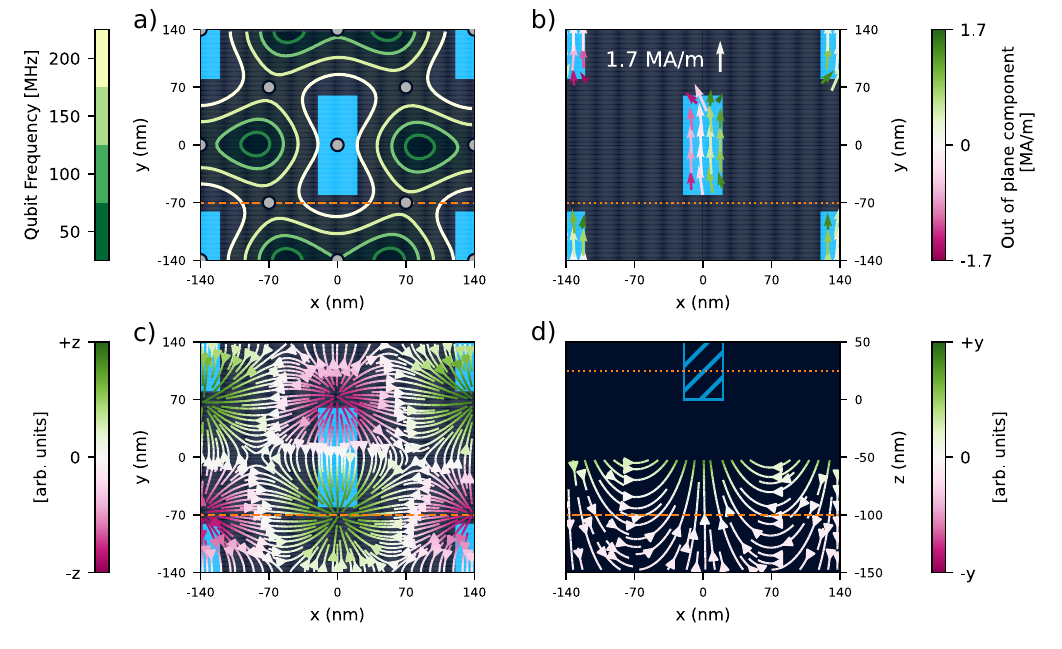}
    \caption{ \textbf{Nanomagnet Simulations} 
    a) Zoomed-in nanomagnet pattern as introduced in Fig.~\ref{Fig:Architecture}. The magnets are shown in bright blue. The bottom surface of the magnets is defined as $z = 0$. For the simulations, we choose Fe as the magnetic material \cite{Aldeghi2023} with saturation magnetization $M_s=\SI{1.7}{\mega\ampere/\meter}$, exchange stiffness $A_{ex} =\SI{21}{\pico\joule/\cubic\meter}$, nanomagnet size $\left\{l_x, l_y, l_z \right\} =\left\{\SI{40}{\nano\meter},\SI{120}{\nano\meter},\SI{50}{\nano\meter}\right\}$, cell size $x,y,z=\SI{5}{\nano\meter}$ and simulated volume $\left\{L_x,L_y,L_z \right\} =\left\{\SI{3990}{\nano\meter},\SI{3815}{\nano\meter},\SI{200}{\nano\meter}\right\}$. The grey dots mark quantum dot locations, and the orange dashed line marks the intersection with the plot shown in panel d). The contour lines show the estimated qubit frequency distribution in a quantum well at $z=\SI{-100}{\nano\meter}$ assuming a $g$-factor of 2 for silicon. No external magnetic field is present.
    b) Magnetization of the nanomagnets at $z = \SI{25}{\nano\meter}$. The dotted orange line marks the intersection with the plot shown in panel d). The color of the quiver arrows shows the z-component of the magnetization. The nanomagnets are originally magnetized along the positive y-axis. After the external field is removed, the shape anisotropy maintains the magnetization predominantly in the y-direction. 
    c) Magnetic field lines in the x-y quantum dot plane at $z = \SI{-100}{\nano\meter}$. The stray field causes the relative quantization axis tips between the dot locations indicated in a) to be about \SI{90}{\deg}. The dashed line indicates the intersection with panel d). 
    d) Stray field lines in the x-z plane at $y = \SI{-70}{\nano\meter}$. The dashed and dotted orange lines mark the intersections with the planes shown a), b)  and c). To maintain good visibility, only the region up to $z = \SI{-50}{\nano\meter}$ is depicted. The crosshatched blue square indicates the position of the magnetic layer, though there is no nanomagnet present in this plane. The inhomogeneity of the stray field extends relatively far below the magnetic layer allowing ample space for independent gate electrodes.}
    \label{Ext_Fig:NanoMag}
\end{figure}
\newpage



\end{document}


\title{Supplementary Information: Baseband control of single-electron silicon spin qubits in two dimensions}

\author{Florian~K.~Unseld}
\thanks{These authors contributed equally to this work.}
\author{Brennan~Undseth}
\thanks{These authors contributed equally to this work.}
\author{Eline~Raymenants}
\author{Yuta~Matsumoto}
\affiliation{QuTech and Kavli Institute of Nanoscience, Delft University of Technology, Lorentzweg 1, 2628 CJ Delft, The Netherlands}
\author{Saurabh~Karwal}
\affiliation{QuTech and Netherlands Organization for Applied Scientific Research (TNO), Stieltjesweg 1, 2628 CK Delft, Netherlands}
\author{Oriol~Pietx-Casas}
\author{Alexander~S.~Ivlev}
\author{Marcel~Meyer}
\affiliation{QuTech and Kavli Institute of Nanoscience, Delft University of Technology, Lorentzweg 1, 2628 CJ Delft, The Netherlands}
\author{Amir~Sammak}
\affiliation{QuTech and Netherlands Organization for Applied Scientific Research (TNO), Stieltjesweg 1, 2628 CK Delft, Netherlands}
\author{Menno~Veldhorst}
\author{Giordano~Scappucci}
\author{Lieven~M.~K.~Vandersypen}
\thanks{L.M.K.Vandersypen@tudelft.nl}
\affiliation{QuTech and Kavli Institute of Nanoscience, Delft University of Technology, Lorentzweg 1, 2628 CJ Delft, The Netherlands}

\maketitle

\onecolumngrid

\renewcommand{\theequation}{S\arabic{equation}}

\renewcommand{\thefigure}{S\arabic{figure}}

This supplementary information includes:

\begin{itemize}
    \item Supplementary Note \ref{supp:EDSR_theory}: Deriving spin properties from magnetic field simulations
    \item Supplementary Note \ref{supp:umagsims}: Micromagnet simulations
    \item Supplementary Note \ref{supp:nanomagsims}: Nanomagnet simulations
    \item Supplementary Note \ref{supp:spinphysics}: Sign extraction from crosstalk experiments
    \item Supplementary Note \ref{supp:hoppingspins}: Hopping spin control and the diabaticity condition
\end{itemize}

\clearpage

\section{Deriving spin properties from magnetic field simulations}
\label{supp:EDSR_theory}

Here, we review how the stray field from on-chip magnets is calculated and how this field determines the spin physics in a 2D array of quantum dots. This is salient as much of the existing literature focuses on the simplified case of linear spin arrays and fails to predict qubit properties away from the sweet spot at low external magnetic field amplitudes. The EDSR Hamiltonian is useful as it elucidates both how a transverse gradient permits qubit control as well as how electric field fluctuations due to charge noise couple to the qubit. Similar derivations may be found in \cite{Pioro_Ladriere_2008} and \cite{Yoneda_2017}. The Hamiltonian has the form:

\begin{equation}
\label{eq:EDSR}
H = -\frac{E_Z}{2}\sigma_z - \frac{\Delta_o}{2}\tau_z + \frac{\Lambda}{2}\tau_x\sigma_x - \frac{\lambda}{2}\tau_x\sigma_z + E(t)\tau_x.
\end{equation}

$\sigma$ and $\tau$ denote Pauli matrices that act on the 2-level spin state or (truncated) 2-level orbital state respectively. $E_Z = g\mu_\textrm{B}B_\textrm{tot}$ is the Zeeman splitting. $g=2$ is the electron spin $g$-factor, $\mu_0$ is the Bohr magneton, $B_{tot}$ is the magnitude of the total magnetic field present at the dot, and $\hbar$ is the reduced Planck's constant. We make the simplifying approximation that the $g$-factor for an electron spin in silicon is isotropic such that the spin's quantization axis is effectively set by the direction of the total magnetic field vector.

$\Delta_o$ is the orbital energy splitting of the dot, which is typically on the order of \SI{1}{\milli\eV}. Although some valley-orbit hybridization may take place, for simplicity we assume it is the orbit-like dipole of length $\langle g|\hat{r}|e\rangle = a_0/\sqrt{2}$ that dominates electric coupling to the charge state of the confined electron. $a_0 = \sqrt{\hbar^2/m^\ast\Delta_0}$ is the length scale of the dot. For the silicon quantum well, $m^\ast$ is 19\% of the free electron mass.

$E(t) = E^\mathrm{drive}(t) + E^\mathrm{noise}(t)$ represents a time-dependent electric field that may consist of an intentional drive as well as environmental charge noise. The electric driving field is given by $E^\mathrm{drive}(t) = eE_\mathrm{ac}^\mathrm{drive}a_0\cos(\omega t)/\sqrt{2}$ and is oriented along the in-plane unit vector $\hat{\mathbf{r}} = r_x\hat{\mathbf{x}} + r_y\hat{\mathbf{y}}$. We assume any out-of-plane component does not participate in the EDSR mechanism. $e$ is the elementary charge of the electron and $E_\mathrm{ac}^\mathrm{drive}\approx\SI{1000}{\volt/\meter}$ is the amplitude of the electric field. Similarly, charge noise fluctuations of a particular frequency $\omega'$ may be expressed as $E^\mathrm{noise}(t) = eE_\mathrm{ac}^\mathrm{noise}a_0\cos(\omega' t)/\sqrt{2}$. The amplitude of the fluctuations $E_\mathrm{ac}^\mathrm{noise}\approx\SI{10}{\volt/\meter}$ may be estimated from charge noise measurements.

$\Lambda = g\mu_\textrm{B}\left|\frac{dB_\perp}{dr}\right|\frac{a_0}{\sqrt{2}}$ gives the energy scale for the synthetic spin-orbit coupling where $\left|\frac{dB_\perp}{dr}\right|$ is the slanting field perpendicular to the quantization axis calculated along the axis of the time-dependent electric fluctuations. The orientation of $B_\perp$ is ambiguous, because there are two orthogonal axes to the quantization axis. This would give rise to $\tau_x\sigma_x$ and $\tau_x\sigma_y$ couplings in the Hamiltonian. By rotating the spin basis about the quantization axis, these contributions can be combined into a single energy scale. $\lambda = g\mu_\textrm{B}\left|\frac{dB_\textrm{tot}}{dr}\right|\frac{a_0}{\sqrt{2}}$ describes the effect of the decoherence gradient, whereby the qubit energy may fluctuate due to charge noise. While we can engineer the orientation of an applied electric drive, charge noise may push the confined electron in any x-y direction, so we aim to calculate a reasonable upper bound of $\left|\frac{dB_\textrm{tot}}{dr}\right|$.

Considering the spin-orbit coupling and time-dependent fluctuation energy scales as small with respect to the orbital splitting, Schrieffer-Wolff perturbation theory may be used to derive the relevant off-diagonal element $\Omega$ for driving spin transitions as well as the diagonal element $\delta\omega_0$ coupling electric fluctuations to the effective qubit energy splitting to first order:

\begin{equation}
\Omega = \frac{g\mu_\textrm{B}a_0^2\left|\frac{dB_\perp}{dr}\right|eE_\mathrm{ac}^\mathrm{drive}}{2\hbar\Delta_o},
\end{equation}

\begin{equation}
\delta\omega_0 = \frac{g\mu_\textrm{B}e\hbar\left|\frac{dB_\textrm{tot}}{dr}\right|E_\mathrm{ac}^\mathrm{noise}}{m^\ast\Delta_o^2}.
\end{equation}

The synthetic spin-orbit coupling also introduces a small, constant renormalization to the Zeeman splitting, but this is unimportant in the context of practical qubit calibration as it is the hybridized qubit frequency $\omega_0$ which is measured directly. These relations directly imply the oft-stated conditions that good EDSR control ($f_\mathrm{Rabi}=\Omega/2\pi > \SI{5}{\mega\hertz}$) is achieved when $\left|\frac{dB_\perp}{dr}\right| > \SI{1}{\milli\tesla/\nano\meter}$ and good charge-noise-limited coherence properties ($T_2^* \approx 1/\delta\omega_0 > \SI{10}{\micro\second}$) should restrict $\left|\frac{dB_\textrm{tot}}{dr}\right| < \SI{0.1}{\milli\tesla/\nano\meter}$. In summary, the total magnetic field, transverse gradient, and decoherence gradient form the most relevant quantities to calculate to predict EDSR behaviour.

In the case of low-field operation when hopping spins may be used for single-qubit gate control, the tip in quantization axes between adjacent quantum dots becomes an important metric. While valley-orbit hybridization and intrinsic spin-orbit coupling will cause some site-to-site variation in the $g$-tensor of the electron spins in the silicon quantum well, the spin quantization axis will predominantly be aligned with the direction of the magnetic field vector at the relevant dot locations \cite{Tanttu_2019}. The tip angle can therefore also be predicted from an accurate magnetic field simulation with quantitative accuracy limited by the accuracy of both the magnet modelling and the electrostatic confinement of the dots.

In Sections~\ref{supp:umagsims} and \ref{supp:nanomagsims}, we summarize the magnetic field simulation strategies employed in this paper. Regardless of how the magnetic vector field is computed, the relevant quantities for understanding spin control can be computed in the same way. We represent the sum of the constant external field and ferromagnetic stray field as a total vector field:

\begin{equation}
\mathbf{B} = \begin{pmatrix} B_x \\ B_y \\ B_z \end{pmatrix}.
\end{equation}

We assume that the qubit locations are point-like as the field varies slowly over the length scale of the electron wavefunction. The total magnetic field $B_\mathrm{tot}$ setting the Zeeman energy is:

\begin{equation}
\label{eq:Btot}
B_\mathrm{tot} = \sqrt{B_x^2 + B_y^2 + B_z^2}.
\end{equation}

With the additional assumption that the $g$-factor of an electron in the conduction band minimum is isotropic, the quantization axis is set by the direction of the total magnetic field:

\begin{equation}
\label{eq:utot}
\hat{\mathbf{u}}_\mathrm{tot} = \frac{B_x}{B_\mathrm{tot}}\hat{\mathbf{x}} + \frac{B_y}{B_\mathrm{tot}}\hat{\mathbf{y}} + \frac{B_z}{B_\mathrm{tot}}\hat{\mathbf{z}}.
\end{equation}

We can find the perpendicular axes using the Gram-Schmidt process, taking $\{\hat{\mathbf{u}}_\textrm{tot}, \hat{\mathbf{x}}, \hat{\mathbf{z}}\}$ as the starting set of normal vectors:

\begin{equation}
\label{eq:gramschmidt}
\hat{\mathbf{u}}_{\perp,1} = \hat{\mathbf{x}} - (\hat{\mathbf{u}}_\textrm{tot}\cdot\hat{\mathbf{x}})\hat{\mathbf{u}}_\textrm{tot}, \\
\hat{\mathbf{u}}_{\perp,2} = \hat{\mathbf{z}} - (\hat{\mathbf{u}}_\textrm{tot}\cdot\hat{\mathbf{z}})\hat{\mathbf{u}}_\textrm{tot} - (\hat{\mathbf{u}}_{\perp,1}\cdot\hat{\mathbf{z}})\hat{\mathbf{u}}_{\perp,1},
\end{equation}

The transverse fields are therefore $B_{\perp,1/2} = \mathbf{B}\cdot\hat{\mathbf{u}}_{\perp,1/2}$ and equal 0 at the qubit locations. The total transverse gradient is found by taking the total directional derivative of each gradient along the driving axis:

\begin{align}
\left|\frac{dB_\perp}{dr}\right| & = \sqrt{(\hat{\mathbf{r}}\cdot\nabla B_{\perp,1})^2 + (\hat{\mathbf{r}}\cdot\nabla B_{\perp,2})^2} \\
& = \sqrt{\left(\frac{\partial B_{\perp,1}}{\partial x}r_x + \frac{\partial B_{\perp,1}}{\partial y}r_y\right)^2 + \left(\frac{\partial B_{\perp,2}}{\partial x}r_x + \frac{\partial B_{\perp,2}}{\partial y}r_y\right)^2}.
\label{eq:driving_gradient}
\end{align}

We refer to Equation~\ref{eq:driving_gradient} as the driving gradient.

For a particular fluctuation axis $\hat{\mathbf{r'}} = r'_x\hat{\mathbf{x}} + r'_y\hat{\mathbf{y}}$, the decoherence gradient is given by the directional derivative:

\begin{equation}
\left|\frac{dB_\textrm{tot}}{dr}\right| = \left|\hat{\mathbf{r'}}\cdot\nabla B_\textrm{tot}\right|.
\end{equation}

Since we don't know along which axis charge noise will predominantly push the dot, we can estimate a bound by taking the norm of the in-plane gradient:

\begin{equation}
\label{eq:decoherence_gradient}
\left|\frac{dB_\textrm{tot}}{dr}\right|_\textrm{max}\approx \sqrt{(\hat{\mathbf{x}}\cdot\nabla B_\textrm{tot})^2 + (\hat{\mathbf{y}}\cdot\nabla B_\textrm{tot})^2}.
\end{equation}

We refer to Equation~\ref{eq:decoherence_gradient} as the decoherence gradient.

Often, the driving axis is designed to align with the same cartesian axis as the external field (e.g. $\hat{\mathbf{y}}$) and the transverse field is dominated by another cartesian component (e.g. $\hat{\mathbf{z}}$) such that the driving gradient can be approximated as $|\partial B_z/\partial y|$. Similarly, the decoherence gradient can be approximated as $|\partial B_y/\partial x| + |\partial B_y/\partial y|$. However, applying these approximations becomes becomes less accurate in regimes where the inhomogeneous stray field of on-chip magnets dominates the uniform external field.

Finally, the quantization axis tip between adjacent spin sites can be computed from the respective magnetic field vectors $\mathbf{B}_1$ and $\mathbf{B}_2$ at each site as:

\begin{equation}
    \label{eq:tip}
    \theta_\mathrm{tip} = \cos^{-1}\left(\frac{\mathbf{B}_1\cdot\mathbf{B}_2}{B_\mathrm{1,tot}B_\mathrm{2,tot}}\right)
\end{equation}

\section{Micromagnet simulations}
\label{supp:umagsims}

For micromagnet simulations where the ferromagnetic material can be approximated as a bulk material, we leverage the efficient magnetic field calculations of the Python package magpylib \cite{magpylib}. Such simulations treat the micromagnet as having a homogeneous polarization with no microscopic crystal structure such that the analytic form of the stray field for constituent magnet shapes can be used to calculate the relevant total magnetic vector field.

There are several sources of uncertainty in such a simulation in our context. At a fundamental level, no time-dynamics or domain wall formation is accounted for in the simulations. See \cite{Aldeghi2024} for a detailed discussion of magnet simulation approaches and their limitations. Independent of the simulation method, the accuracy is limited to the level of detail included in the magnet model itself as well as knowledge of the precise location of the accumulated quantum dots.

We therefore aim to extract a useful qualitative picture. First, we model the micromagnet using the nominal design. This excludes roughness due to the underlying gate structure, finite rounding of the magnet edges, and other small misalignments from the fabrication procedure. Second, we use the experimentally measured qubit frequencies to fit the effective homogeneous polarization of the model at a particular external field setting using a least-squares optimization (see Fig.~3). This requires making an assumption about the location of the spins. We take the center coordinate of the plunger gates as the point-like dot location (as indicated by the red dots in the bottom row of External Data Fig.~9). A more detailed electrostatic simulation could be leveraged to predict the dot locations with more precision, but it is unclear if this confers any benefit due to the approximate nature of the magnetic simulation itself. With a fitted magnet polarization, a field can be simulated from which relevant parameters may be extracted per Section~\ref{supp:EDSR_theory}.

The trends predicted in terms of addressability, driving gradient and decoherence gradient all qualitatively match our experimental observations. While the model also predicts when quantization axis tips become substantial, a quantitative estimate becomes very sensitive to the estimated dot locations and the microscopic demagnetization of the micromagnets. Leveraging hopping spins through a larger array could be a powerful \textit{in situ} method of characterizing micromagnet behaviour. Direct cryogenic magnetic imaging of patterned micromagnets would be another straightforward means of verifying magnet behavior.

\section{Nanomagnet simulations}
\label{supp:nanomagsims}

To obtain an accurate picture of the periodic stray fields produced by the nanomagnets in the 2D array, we use the OOMMF package \cite{OOMMF}. We set the nanomagnets’ initial magnetization along the y-axis (their longest axis), which we would perform experimentally by applying an external B-field along the y-axis. We employ an energy minimization evolver/driver (MinDriver) to find the energy-relaxed state of the entire 2D array.  The relaxed magnetization will be defined by the nanomagnets’ material parameters, as well as their geometry (shape anisotropy), pitch and initial magnetization direction. For the results presented in the main text, we choose Fe with material parameters similar to those in \cite{Aldeghi2023}, i.e. saturation magnetization $M_s$ = 1700 kA/m, exchange stiffness $A_{ex}$ = 21 pJ/m, and \SI{0}{\kelvin} temperature. We aim for a quantum dot pitch of 100 nm, which requires a nanomagnet pitch of about 280 nm in both the vertical and horizontal direction (this is the pitch within the same row or column). To minimize edge effects, we simulate a rather large volume of 3990 nm x 3815 nm x 200 nm with a cell size of 5 nm in all directions. The nanomagnet sizes are 40 nm x 120 nm x 50 nm. The results in the main text present stray fields at a distance of 100 nm below the nanomagnets, mimicking a quantum well positioned 100 nm below the nanomagnet array. Additional simulations suggest that the exact distance is not critical, as favorable conditions for hopping occur at a range of 90 to 140 nm for the specific sizes of the Fe nanomagnets shown here. Moreover, we find that by decreasing the thickness of the Fe nanomagnet to \SI{30}{\nano\meter}, we can decrease the qubit frequency and decoherence gradient further while maintaining the same tip angles. Finally, we note that we are not limited to Fe, as we further varied ferromagnetic materials and nanomagnet geometries and could obtain similar outcomes. We leave these optimizations for future studies.

\section{Sign extraction from crosstalk experiments}
\label{supp:spinphysics}

We make use of EDSR to characterize spin physics and encode a set of universal gates with which to perform state tomography. Using the convention $g > 0$ for the electron spin in silicon, the effective two-level Hamiltonian may be given as:

\begin{equation}
    \label{eq:ESR_spin_MW}
    H = \omega_0 S_z + 2\Omega\cos(\omega_\mathrm{MW}t+\phi)S_x,
\end{equation}

\noindent where $\omega_0 \approx g\mu_0 B_{tot}/\hbar$ is the measured qubit Larmor frequency and $\Omega$ is the Rabi frequency due to the effective ac magnetic drive derived in Section~\ref{supp:EDSR_theory}. Here, a microwave drive of frequency $\omega_\mathrm{MW}$ and phase $\phi$ are used for EDSR. The Loss-DiVincenzo qubit operators are encoded by the spin operators as $\sigma_z = -2S_z/\hbar$ such that $\ket{\downarrow} = \ket{0}$ and $\ket{\uparrow}=\ket{1}$. Therefore, now taking $\hbar=1$, Equation~\ref{eq:ESR_spin_MW} may be rewritten as:

\begin{equation}
    \label{eq:EDSR_bloch}
    H = -\frac{\omega_0}{2}\sigma_z + \Omega\cos(\omega_\mathrm{MW}t+\phi)\sigma_x.
\end{equation}

Experiments at microwave frequencies make use of the rotating frame where $\ket{\tilde{\psi}(t)} = R_z(\omega_\mathrm{MW}t)\ket{\psi(t)}$. In our notation, the unitary operator $R_n(\alpha)=\exp(-i\alpha\hat{n}\cdot\vec{\sigma}/2)$ represents a positive rotation in the Bloch sphere about the unit vector $\hat{n} = (n_x,n_y,n_z)$ by an angle $\alpha$. The rotating frame Hamiltonian, neglecting fast oscillating terms, is given by:

\begin{align}
    \label{eq:EDSR_rot}
    \tilde{H} & = R_z(\omega_\mathrm{MW}t)HR_z^\dagger(\omega_\mathrm{MW}t) + i \frac{dR_z(\omega_\mathrm{MW}t)}{dt}R_z^\dagger(\omega_\mathrm{MW}t) \\
    & = \frac{\Delta}{2}\sigma_z + \frac{\Omega}{2}(\cos\phi\sigma_x-\sin\phi\sigma_y),
\end{align}

\noindent where $\Delta = \omega_\mathrm{MW}-\omega_0$. During free evolution in the lab frame for a time $t$, $H_\mathrm{free} = -\frac{\omega_0}{2}\sigma_z$, and the state evolves with $U_\mathrm{free}(t) = R_z(-\omega_0 t)$. During free evolution in the rotating frame, $\tilde{H}_\mathrm{free} = \frac{\Delta}{2}\sigma_z$, and the state evolves with $\tilde{U}_\mathrm{free}(t) = R_z(\Delta t)$.

We use parity-mode Pauli spin blockade for measurement and take our observable as $O=(\mathbb{1} - \sigma_z\otimes\sigma_z)/2$ in all cases where $\mathbb{1}$ is the identity operator. We initialize using post-selection into the subspace spanned by $\{\ket{01},\ket{10}\}$ (see Extended Data Fig.~8 for further discussion). Regardless of whether the initial pair is entangled, mixed, or a product state, the following expectation values hold.

We can use a modified Ramsey sequence with a virtual detuning in order to precisely estimate the qubit frequency. After preparation of a superposition state via a gate applied to qubit $i$, an $R_z^i(\omega_\mathrm{virtual}t)$ operation is performed via a reference frame update (in the case of resonant control) or a physical wait (in the case of baseband control) prior to measurement:

\begin{equation}    \mathrm{Init}\rightarrow R_x^i(\pi/2)\rightarrow\mathrm{Wait}~t\rightarrow R_z^i(\omega_\mathrm{virtual}t)\rightarrow R_x^i(-\pi/2)\rightarrow\mathrm{Measure}~O.
\end{equation}

\noindent The expectation value of the time-ordered sequence is:

\begin{equation}
p_\mathrm{odd}(t) = A\cos((\Delta+\omega_\mathrm{virtual})t)\exp(-(t/T_2^*)^a)+B,
\end{equation}

\noindent where $p_\mathrm{odd}(t)$ is the probability of measuring a state in the $\{\ket{01},\ket{10}\}$ subspace. $A$ and $B$ are visibility and offset corrections due to constant state preparation and measurement (SPAM) errors, $a$ is a decay constant and $T_2^*$ is the decay time. The frequency of the fitted oscillations $\omega_\mathrm{fit}^*$ can be related to the Larmor frequency as:

\begin{equation}
    \label{eq:ramsey_rot_fit}
    \omega_0 = \omega_\mathrm{MW}-(\omega_\mathrm{fit}^*-\omega_\mathrm{virtual}).
\end{equation}

A similar analysis holds for a Hahn echo pulse sequence:

\begin{equation}    \mathrm{Init}\rightarrow R_x^i(\pi/2)\rightarrow\mathrm{Wait}~t/2\rightarrow R_x^i(\pi)\rightarrow \mathrm{Wait}~t/2\rightarrow R_z^i(\omega_\mathrm{virtual}t)\rightarrow R_x^i(-\pi/2)\rightarrow\mathrm{Measure}~O.
\end{equation}

\noindent The expectation value is given as:

\begin{equation}
p_\mathrm{odd}(t) = C\cos((\Delta_2-\Delta_1)/2+\omega_\mathrm{virtual})t)\exp(-(t/T_2^\mathrm{H})^b)+D,
\end{equation}

\noindent where $C$ and $D$ are once again due to SPAM errors, $b$ is a decay constant and $T_2^H$ is the extracted decay time. Here we observe that, contrary to expectation, the free evolution in the rotating frame with frequency $\Delta_1=\omega_\mathrm{MW}-\omega_{01}$ before the decoupling pulse may be measurably different than the free evolution with frequency $\Delta_2=\omega_\mathrm{MW}-\omega_{02}$ after the pulse. Fitting the decaying oscillations to a frequency $\omega_\mathrm{fit}^H$ offers information about this systematic difference in Larmor frequency after the echo pulse, as any quasistatic fluctuations are eliminated by the echo pulse:

\begin{equation}
    \label{eq:hahn_rot_fit}
    \omega_{01}-\omega_{02} = 2(\omega_\mathrm{fit}^H-\omega_\mathrm{virtual}).
\end{equation}

An off-resonant pulse may be embedded into the second half of an echo sequence of fixed duration $T$ to probe the transient phase pickup $\theta_t(t_\mathrm{delay})$ over time $t_\mathrm{delay}$ as in Fig.~2. In this case, a gate $R_z^i(\theta)$ is swept prior to measurement:

\begin{multline}    \mathrm{Init}\rightarrow R_x^i(\pi/2)\rightarrow\mathrm{Wait}~T/2\rightarrow R_x^i(\pi)\rightarrow \mathrm{Wait}~T/2-t_\mathrm{delay}\rightarrow\mathrm{Burst}\rightarrow \\
\mathrm{Wait}~t_\mathrm{delay}\rightarrow R_z^i(\theta)\rightarrow R_x^i(\pi/2)\rightarrow\mathrm{Measure}~O.
\end{multline}

\noindent The expectation value is given as:

\begin{equation}
\label{eq:PIRS_fit}
p_\mathrm{odd}(\theta,t_\mathrm{delay}) = E\cos(T(\Delta_2-\Delta_1)/2+\theta_t(t_\mathrm{delay})+\theta) + F,
\end{equation}

\noindent where $E$ and $F$ account for the loss of visibility due to the echo decay as well as SPAM errors. As the difference $\Delta_2-\Delta_1$ is known from the unmodified echo experiment, fitting the oscillation permits extracting $\theta_t$. In the rotating frame, $\theta_t = \int dt_\mathrm{delay}\Delta(t_\mathrm{delay})$, and it follows that $\omega_0(t_\mathrm{delay}) = \omega_\mathrm{MW} - d\theta_t/dt_\mathrm{delay}$.

\section{Hopping spin control and the diabaticity condition}
\label{supp:hoppingspins}

\begin{figure}[h!]
    \centering
    \includegraphics[width=.9\textwidth]{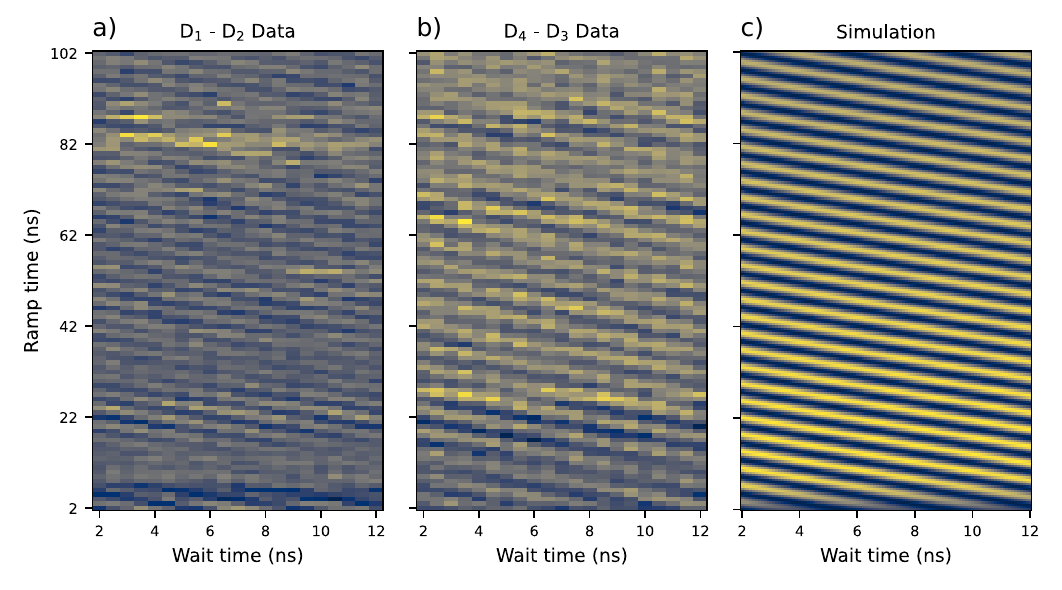}
    \caption{ \textbf{Spin shuttling versus ramp time} a) Observed spin fraction oscillations when shuttling between Dots 1 and 2 with a varying ramp time. b) Observed spin fraction oscillations when shuttling between Dots 4 and 3 with a varying ramp time. c) Simulation using parameters extracted from Pair 1-2 in experiment. A total detuning ramp of \SI{300}{\milli\volt} is converted to \SI{10500}{\micro\eV} with a combined lever arm of \SI{0.07}{\milli\eV/\milli\volt}, and a tunnel coupling of \SI{30}{\micro\electronvolt} is estimated. The measured qubit frequencies of \SI{240}{\mega\hertz} and \SI{270}{\mega\hertz} are used for the two dot positions along with a measured tip angle of \SI{22}{\deg}. The data in a-b) was collected with an external field of \SI{-15}{\milli\tesla}. The simulations show that the sudden approximation for spin-state preservation during shuttling is reasonably met in our experiments over a large range of ramp times owing to the very large detuning sweep.}
    \label{Supp_Fig:ramptime}
\end{figure}

Single-spin control via hopping spins can be understood starting from a Hamiltonian describing a single electron in a double quantum dot:

\begin{align}
    \label{eq:HDQD}
    H_\mathrm{DQD} & = H_\mathrm{charge} + H_\mathrm{spin} \nonumber \\  
    & = \begin{pmatrix}
        \epsilon/2 & 0 & t_c & 0 \\
        0 & \epsilon/2 & 0 & t_c \\
        t_c & 0 & -\epsilon/2 & 0 \\
        0 & t_c & 0 & -\epsilon/2 \\
    \end{pmatrix} + \frac{g\mu_B}{2}
    \begin{pmatrix}
        B_\mathrm{1,tot} & 0 & 0 & 0 \\
        0 & -B_\mathrm{1,tot} & 0 & 0 \\
        0 & 0 & B_\mathrm{2,tot}\cos(\theta_\mathrm{tip}) & B_\mathrm{2,tot}\sin(\theta_\mathrm{tip})e^{-i\phi} \\
        0 & 0 & B_\mathrm{2,tot}\sin(\theta_\mathrm{tip})e^{i\phi} & -B_\mathrm{2,tot}\cos(\theta_\mathrm{tip})
    \end{pmatrix},
\end{align}

\noindent where $\epsilon$ and $t_c$ are the detuning and tunnel coupling describing the double dot system. In the second term, the Zeeman energies at the two sites are considered where $\theta_\mathrm{tip}$ is the polar angle between the two quantization axes and $\phi$ is the azimuthal angle. This model is directly analogous to the one employed in \cite{van_Riggelen_Doelman_2024, Wang2024} with the difference that the tip in quantization is attributed to the magnetic field as opposed to the $g$-tensor. As we are only interested in single-spin physics, we may take $\phi=0$ with an appropriate choice of coordinate frame.

Fig.~4b) plots the eigenenergies of Equation~\ref{eq:HDQD}. When the detuning is linearly swept $\epsilon(t) = vt$, the Landau-Zener formula provides a means to estimate the condition for which charge transfer between the two dots is is not adiabatic:

\begin{equation}
    \label{eq:LZ}
    P_\mathrm{LZ} = \exp\left(\frac{-2\pi t_c^2}{\hbar v}\right).
\end{equation}

\noindent Based on a combined vP1+vP2 lever arm of about \SI{0.07}{\milli\eV/\milli\volt} and a ramp time of \SI{5}{\nano\second}, we estimate a detuning ramp speed of about $v=\SI{3000}{\micro\eV/\nano\second}$. With an estimated tunnel coupling of \SI{40}{\micro\eV}, we roughly estimate that $0.001 < P_\mathrm{LZ}\approx 0.01$. We discuss the implications of this shortly.

Assuming the electron charge is transferred adiabatically, the resulting spin physics is clarified by transforming to the diagonal basis of $H_\mathrm{charge}$ with a unitary $U = \exp\left(-i\tan^{-1}\left(-2t_c/\epsilon\right)\sigma_y\otimes\sigma_0\right)$ and investigating $H' = UH_\mathrm{spin}U^\dagger$ in the sector corresponding to a ground charge state:

\begin{multline}
    \label{eq:Hspindiabatic}
    H' = \frac{g\mu_B}{4}\Bigg[\Bigg.\left(B_\mathrm{1,tot}\left(1-\frac{\epsilon}{\sqrt{\epsilon^2+4t_c^2}}\right)+B_\mathrm{2,tot}\cos\theta_\mathrm{tip}\left(1+\frac{\epsilon}{\sqrt{\epsilon^2+4t_c^2}}\right)\right)\sigma_z \\
    +\left(B_\mathrm{2,tot}\sin\theta_\mathrm{tip}\left(1+\frac{\epsilon}{\sqrt{\epsilon^2+4t_c^2}}\right)\right)\sigma_x\Bigg.\Bigg]
\end{multline}

Fig.~4c) plots the components of Equation~\ref{eq:Hspindiabatic}. A clear detuning-dependent step arises that is not evident from inspecting the energy level diagram alone. The sudden approximation quantifies how quickly such a Hamiltonian step needs to be traversed in order for the initial quantum state to be preserved:

\begin{equation}
    \label{eq:suddenapproximation}
    t\ll \frac{2\pi\hbar}{\Delta E},
\end{equation}

\noindent where $\Delta E\approx\SI{0.1}{\micro\eV}$ is the difference between the relevant eigenvalues of $H'$. This suggests the time interval needs to be less than \SI{1}{\nano\second}. As the energetic width of the step is about $4t_c$ for the case of our linear detuning ramp, the step is crossed in about \SI{60}{\pico\second}. In fact, even a total ramp time much longer than was used for the logical gate implementation may be sufficient. This is observed in Fig.~\ref{Supp_Fig:ramptime} where ramp times extending to several tens of nanoseconds were used while still observing evidence of sudden spin state transfer between both dot pairs 1-2 and 3-4. Simulations of the time-dependent Schrodinger equation with estimates of experimental parameters corroborate this.

The estimated 0.1-1\% probability of a diabatic transition at the charge anticrossing is likely the dominant limitation for the maximum single-qubit gate fidelity that we achieved via hopping spins. This was due to a particularly limited window of device stability within which DC voltages could be tuned to either increase the tunnel coupling or decrease the detuning range necessary to shuttle the electron. Equation~\ref{eq:LZ} and Equation~\ref{eq:suddenapproximation} represent a fundamental trade-off, as we desire an adiabatic charge transition with a diabatic spin transition. However, the exponential dependence of the Landau-Zener formula implies that even a modest increase of tunnel coupling greatly enhances the quality of charge transfer while still safely satisfying the sudden approximation. Furthermore, the increased susceptibility to charge noise that occurs when the spin and charge are maximally hybridized means that a fast shuttling time is preferable.

As the micromagnet design and gate layout in the present device was not intended for electron shuttling, there is almost certainly substantial room for improved fidelities by tailoring device design to the conditions for performing shuttling gates. Building upon the understanding of the depolarization of the micromagnet under low-field conditions, engineering quantum axis tips well above 45 degrees between adjacent dots should be easily achievable while keeping qubit frequencies on the order of \SI{100}{\mega\hertz}. Furthermore, being able to modulate the tunnel coupling to the range of \SI{50}{}-\SI{100}{\micro\eV} would substantially improve the quality of adiabatic charge shuttling and should lead to enhanced gate operation.

\bibliography{sn-bibliography}